\documentclass[11pt,a4paper]{article}
\pdfoutput=1
\usepackage{jcappub}
\usepackage[T1]{fontenc} % if needed
\usepackage{lmodern}
\usepackage{xcolor,graphicx}
\usepackage{caption}
\usepackage{subcaption}
\usepackage{amsmath,amssymb}
\usepackage[font=small,labelfont=bf]{caption}
\usepackage{verbatim}
\usepackage{placeins}
\usepackage{tikz}
\usetikzlibrary{arrows, positioning, angles,quotes, calc, intersections}
\usepackage{enumitem}
\def\mean#1{\left< #1 \right>}

\usepackage{color}

\usepackage[normalem]{ulem}
\usepackage{enumitem}

\newcommand{\bk}{{\mathbf k}}

\newcommand{\bB}{{\mathbf B}}

\newcommand{\bx}{{\mathbf x}}

\newcommand{\bv}{{\mathbf v}}
\newcommand{\bnabla}{{\boldsymbol\nabla}}
\newcommand{\bom}{{\boldsymbol\omega}}

\newcommand{\de}{\delta}
\newcommand{\De}{\Delta}

\newcommand{\ga}{\gamma}

\newcommand{\Om}{\Omega}
\newcommand{\om}{\omega}
\newcommand{\si}{\sigma}

\newcommand{\be}{\begin{equation}}
\newcommand{\ee}{\end{equation}}

\newcommand{\bea}{\begin{eqnarray}}
\newcommand{\eea}{\end{eqnarray}}
\newcommand{\bean}{\begin{eqnarray*}}
\newcommand{\eean}{\end{eqnarray*}}

\newcommand{\HH}{{\cal H}}

\newcommand{\inspire}[1]{\href{http://inspirehep.net/search?p=find+J+#1}
 {{\color{black}[{\color{blue} {\small in}SPIRE}]}}}
\newcommand{\book}[1]{\href{http://inspirehep.net/search?p=#1}
 {{\color{black}[{\color{blue} {\small in}SPIRE}]}}}

\newcommand{\inspired}[1]{\href{http://inspirehep.net/search?p=#1}
 {{\color{black}[{\color{blue} {\small in}SPIRE}]}}}

\usepackage{sidecap}
\sidecaptionvpos{figure}{c}

\title{The Generation of Vorticity \\ in Cosmological N-body Simulations }

\author[a]{Goran Jelic-Cizmek}
\author[b,c]{\!, Francesca Lepori}
\author[d]{\!, Julian Adamek}
\author[a]{and Ruth Durrer}

\affiliation[a]{Universit\'e de Gen\`eve, D\'epartement de Physique Th\'eorique and Centre for Astroparticle Physics,
24 quai Ernest-Ansermet, CH-1211 Gen\`eve 4, Switzerland}
\affiliation[b]{SISSA -- International School for Advanced Studies, Via Bonomea 265, 34136 Trieste, Italy}
\affiliation[c]{INFN, Sezione di Trieste, Via Valerio 2, I-34127 Trieste, Italy}
\affiliation[d]{School of Physics \& Astronomy, Queen Mary University of London, 327 Mile End Road, London E1 4NS, United Kingdom}

\emailAdd{goran.jelic-cizmek@unige.ch}
\emailAdd{flepori@sissa.it}
\emailAdd{julian.adamek@qmul.ac.uk}
\emailAdd{ruth.durrer@unige.ch}

\abstract{Clustering of a perfect fluid does not lead to the generation of vorticity. It is the collisionless nature of dark matter, inducing velocity dispersion and shell crossing, which is at the origin of cosmological vorticity generation.
In this paper we investigate the generation of vorticity during the formation of cosmological large scale structure using the public relativistic N-body code \emph{gevolution}. We test several methods to compute the vorticity power spectrum and we study its convergence with respect to the mass and grid resolution of our simulations. We determine the power spectrum, the spectral index on large-scales, the amplitude of the peak position and their time evolution. We also compare the vorticity extracted from our simulations with the vector perturbations of the metric. Our results are accompanied by resolution studies and compared with previous studies in the literature.
}

\begin{document}

\maketitle

\section{Introduction and motivation}
\label{s:intro}
It is well known that most galaxies rotate and that the angular velocities of neighboring galaxies are correlated. It has recently been argued that observations of radio galaxies indicate that these correlations extend also to substantially larger scales, up to 10-20 Mpc~\cite{Taylor:2016rsd}. While we only observe baryons, in most physical dark matter models one expects that similar correlations exist between the respective host dark matter halos.
Within linear perturbation theory, cold dark matter (CDM) is usually modelled as a pressureless perfect fluid. However, since vorticity cannot be generated in a pressureless perfect fluid that is initially free of vorticity, and since within standard cosmology the initial vorticity is negligible, this vortical motion must be modelled with theories going beyond the perfect fluid description of the cosmological matter.

The evolution of small (linear) perturbations in the early Universe is such that the fluid rotational velocity $\Om$ decays like $1/a$ in a matter dominated Universe~\cite{RuthBook}, where $a$ denotes the cosmic scale factor. Even though the amplitude of $\Om$ remains constant in a radiation dominated Universe, it must be very small initially since it is of the order of $\Omega \simeq (k/\HH)^2\si^{(V)}$, where $\si^{(V)}$ is the helicity-1 (vortical) contribution to the shear, which can be at most of the order of $10^{-4}$ after inflation and the factor $(k/\HH)$ is very small after inflation for all cosmologically relevant scales. A rough order of magnitude for the value of the Hubble parameter after reheating is $\HH_{\rm RH}\sim 10^{10}(T_{\rm RH}/1{\text{ TeV}})$Mpc$^{-1}$. For a cosmological scale of 1 Mpc this yields $(k/\HH_{\rm RH})^2\lesssim 10^{-20}$, hence this possible perturbative vorticity can safely be neglected.
This is under the assumption that vorticity is generated by some non-standard inflationary model. Most inflationary models actually do not generate helicity-1 modes to begin with, so that $\si^{(V)}=0$ after inflation.

Furthermore, according to Helmholtz's theorem, a pressureless perfect fluid that is initially irrotational remains irrotational in the absence of external rotational forces also beyond perturbation theory~\cite{Helmholtz1858}\footnote{Helmholtz's original work assumed incompressibility for the fluid. However, his law was 
generalized beyond this approximation by other authors. The validity of the theorem for a barotropic fluid was proved by Hankel (see \cite{Villone2017, Frisch2017} for an english translation of his work).}. This theorem is also non-perturbatively valid in General Relativity~\cite{Lu:2008ju}. In order to generate vorticity, a non-vanishing velocity dispersion is required. 
Assuming velocity dispersion to be present in dark matter at some given level, the induced vorticity has been determined perturbatively in~\cite{Cusin:2016zvu}. 
In this approach, velocity dispersion is an intrinsic property of the dark matter fluid, while the velocity dispersion driven by the shell-crossing of cold dark matter particles cannot be encoded in this framework. Furthermore, it has the drawback to require the introduction of a free parameter for the velocity dispersion which regulates the amplitude of the vorticity spectrum.
Several studies of shell-crossing in the context of perturbation theory have been recently presented (see \cite{Rampf:2017jan, McDonald:2017ths, Pietroni:2018ebj, Saga:2018nud} and references therein). 
However, the application of these techniques to the full 3D problem is still lacking.

The alternative to a perturbative approach is to numerically investigate the generation of vorticity.
Here we pursue this approach.
The dark matter phase space distribution obeys the BBGKY hierarchy which for weak long range forces is well approximated as a Vlasov-Poisson system \cite{Bernardeau:2001qr}.
Therefore, a full numerical study would require solving the full system for the six-dimensional phase space.
From a computational point of view, this is a very challenging task.
There have been several attempts in the literature to solve the Vlasov-Poisson system for the full phase space distribution function (see \cite{Sousbie:2015, Widrow:1993qq, Kopp:2017hbb, Mocz:2018ium} and references therein).
Nevertheless, nowadays the most common approach is to simplify the problem by resorting to the N-body method, which samples the phase space distribution function at some discrete locations corresponding to the particle positions and velocities.
N-body simulations are an extremely powerful tool to study the large-scale structure of the Universe in the non-linear regime, capturing very complex astrophysical phenomena \cite{1985ApJS...57..241E,2000astro.ph..7362F,Springel:2005nw, Springel:2006vs, BoylanKolchin:2009nc, Shandarin:2011jv, Springel:2009aa, Springel:2017tpz} and including gravitational effects beyond the Newtonian approximation \cite{Adamek:2014xba, Adamek:2015eda, Adamek:2016zes, Adamek:2017uiq}.
In cosmological N-body simulations, structure formation leads to shell (orbit) crossing, i.e. particles at (nearly) the same position can have very different velocities. This induces significant velocity dispersion which in turn sources vorticity. The fact that N-body simulations have a finite mass and spatial resolution induces additional (numerical) velocity dispersion and it is crucial to carefully check whether we can distinguish the latter numerical artifact from the former physical phenomenon. 

A first numerical study of the vorticity field with N-body simulations was presented in \cite{Pueblas:2008uv}, where the vorticity power spectrum has been computed in simulations with different
mass resolutions, thus raising the issue of the resolution dependence of the vorticity spectrum in N-body simulations. The vorticity power spectrum was computed from N-body simulations also in \cite{Thomas:2015kua} and~\cite{Hahn:2014lca}, where it was also studied in the context of the warm dark matter scenario.

In this work we shall perform N-body simulations using the publicly available relativistic code \emph{gevolution}~\cite{Adamek:2015eda,Adamek:2016zes}. 
Here the relativistic aspect of the code is less important than the fact that we work with a fixed grid, which also determines our spatial resolution. In the previous studies the spatial resolution was instead provided by the softening length, a numerical artifact used in the computation of the forces to prevent divergences.
Even though the results of the previous studies qualitatively agree, \cite{Pueblas:2008uv} and~\cite{Thomas:2015kua} find e.g. a different time evolution for the vorticity at relatively high redshift ($z > 1$).
Apart from the different N-body code, in this paper we use a different method to compute the vorticity and we compare our results with these previous studies.
The goal of this work is to improve our understanding of the vorticity generation process in N-body simulations and its spectral properties, which is crucial in order to properly model the effect of vector perturbations in redshift-space clustering analysis \cite{Zhu:2017vtj, Bonvin:2017req} and which may be useful for understanding the intrinsic alignments of galaxies \cite{Troxel:2014dba}.

The remainder of this paper is organized as follows. In the next section we detail our method, define the relevant quantities and we describe the numerical calculations and the tests we perform. In section~\ref{s:res} we present our results and compare them with previous studies. In section~\ref{s:con} we sum up and give a conclusion.
In two appendices we study resolution effects and compare our method with the Delaunay tessellation method used in~\cite{Pueblas:2008uv}.
\vspace{0.2cm}

{\bf Notation:} We use the metric in the Poisson gauge $ds^2 = a^2[-(1+2\Psi)dt^2-2B_i dx^i dt+(1-2\Phi)\de_{ij}dx^idx^j ]$, hence $t$ denotes conformal time. Here $B^i_{,i}=0$, hence $B_i$ denotes a vector perturbation, 'frame dragging' and we neglect tensor perturbations (gravitational waves). While 4d indices are denoted by greek letters and raised and lowered with the metric $g_{\mu\nu}$, spatial, 3d indices are denoted by latin letters. They are raised and lowered with the flat 3d metric $\de_{ij}$ and hence 3d index positions are irrelevant. The conformal Hubble parameter is denoted by $\HH=\dot a/a = aH$, where an overdot denotes the derivative with respect to conformal time $t$. Spatial indices are denoted by Latin letters, and spacetime indices by Greek letters. Spatial vectors are denoted in bold face.
In the figures, the velocity divergence and vorticity are normalized in such a way that the linear velocity divergence power spectrum matches the linear matter density power spectrum. 

\section{Method}
\label{s:method}
In this section we will describe the numerical implementation of the vorticity field in the N-body code \emph{gevolution}~\cite{gevolution-web}.

The velocity field can be split into a gradient and the rotational part such that
\be
\bv = -\bnabla v + \bv_R \equiv \bv_G + \bv_R~\,, \mbox{ with }\qquad   \bnabla \cdot \bv_R =0 \,.
\ee
We define the divergence $\theta$ and the vorticity $\boldsymbol{\om}$ by
\bea
\theta &=&  \bnabla \cdot \bv = -\De v \,, \\
\bom &=&\bnabla \wedge\bv = \bnabla \wedge\bv_{R}. 
\eea  

The implementation of the vorticity is performed in two steps: the computation of the velocity field in real space that is detailed in section \ref{s:method1}, and the projection of the velocity in its
divergence and rotational part which is described in section \ref{s:method2}.

The linear power spectrum displayed in our plot for comparison and the linear transfer functions needed for the initial conditions of our simulations are computed with the Cosmic Linear Anisotropy Solving System ({\sc class}) code~\cite{class1, class2}. 
We assume a spatially flat $\Lambda$CDM cosmology compatible with the Planck 2015 results~\cite{Planck2015}: $h = 0.67556$, $\Omega_\text{cdm} h^2 = 0.12038$, $\Omega_\text{b}h^2 = 0.022032$.
The primordial amplitude and spectral index are $A_\text{s} = 2.215 \times 10^{-9}$ and $n_\text{s} = 0.9619$, respectively, at the pivot scale $k_*=0.05\,\text{Mpc}^{-1}$.
Note that the initial velocity field in linear perturbation theory is a pure gradient, and the particles are assigned the velocities only after their initial displacement $x_i \rightarrow x'_i=x_i + \delta x_i(\mathbf{x})$, that is, the velocity field is evaluated at the displaced positions, ensuring that initially we have $\bom = 0$.
Due to finite numerical precision and CIC projection effects, however, some spurious vorticity is still generated initially.
The latter is resolution dependent, and in section \ref{s:converg} we study the convergence of the obtained vorticity power spectrum with respect to the resolution of the simulation.

\subsection{Computation of the velocity field}
\label{s:method1}
In $\textit{gevolution}$, particles are evolved using the geodesic equation and their energy momentum tensor is then calculated by a particle-to-mesh projection, see~\cite{Adamek:2015eda,Adamek:2016zes} for details.
The stress-energy tensor components are related to the density and the momentum density as follows:
\be
T^0_0(\bx) = -\rho(\bx), \qquad T^i_0(\bx) = -\rho(\bx) v^i(\bx), \qquad T^i_j(\bx) = \rho(\bx)\, v^i(\bx) v_j(\bx). 
\ee

We define the velocity field simply by
\begin{equation}
 v^i(\bx) = \frac{T^i_0(\bx)}{T^0_0(\bx)}. \label{vdef}
\end{equation}

The particle-to-mesh projection for both the $T^0_0$ and $T^i_0$ components adopts the Cloud-In-Cell (CIC) method \cite{Hockney:1988:CSU:62815}.
Details about the implementation of the CIC method in \textit{gevolution} can be found in appendix B of \cite{Adamek:2016zes} or in the monograph on N-body simulations~\cite{SimBook}. 
We use the CIC projection also for the momentum density, in order to obtain a velocity that is effectively a mass-weighted average in each cell.
In our simulations, vector metric perturbations are included in the equations of motion, while in the computation of the velocity field we neglect them.
The definition \eqref{vdef} is singular if the cells surrounding the grid point $\bx$ are empty and thus $\rho(\bx)=0$, i.e. inside voids. 
This is of course a consequence of the finite mass resolution of our N-body simulations and it raises the well-known issue of characterizing the velocity field from a discrete, and not uniformly distributed, sample of points.
In the literature there are several ways to approach this problem. In \cite{Bertschinger1990} a smoothing is employed to extract the velocity field from a galaxy redshift survey\footnote{The application to a galaxy catalogue presents the further complication that the galaxies' velocity vectors are known only in the line-of-sight direction.}. In this approach, the discrete velocity field is convolved with a kernel with non-compact
support. Therefore, the smoothed density field is never identically zero. Another approach which has been successfully applied to N-body simulations, consists of building a tessellation from the irregular distribution of particles \cite{Bernardeau:1995en}. In particular, the Delaunay tessellation has been used to extract the velocity and vorticity field from N-body simulations in \cite{Pueblas:2008uv, Thomas:2015kua}. A refined method, which employs a phase-space interpolation technique, was presented in \cite{Abel2012, Hahn:2014lca}. This method appears to perform better than the tessellation method, especially on small scales. 
However, both the phase space interpolation and the tessellation results are difficult to incorporate in a pre-built N-body code, and they usually require to post-process a snapshot of the simulation.
In this work we tested four alternative methods that allow us to compute the velocity field on the fly and, therefore, reduce the computational cost and the memory requirement.

The first three methods estimate the velocity field from Eq.\ \eqref{vdef} in the domain of the simulation where the density is not zero,
while in the voids we invoke the following three mechanisms to assign velocities:
 
\begin{enumerate}
\item We set $\bv=0$ to the grid points belonging to an empty cell. This is the simplest prescription, however it neglects the expected outflows of the particles from low to high density regions. We call this the {\it zero} method.
\item In the initial conditions all grid points have a finite mass density, $\rho(\bx,t_{\rm in})\neq0$. When $\rho(\bx,t_i)=0$ in some grid point $\bx$, we set $v(x,t_{i})=v(x,t_{i-1}) $. 
According to this prescription, the velocity field is frozen when a void region forms in the simulation. We call this the {\it past} method.
\item When $\rho(\bx,t_i)=0$ in some grid point $\bx$, we set $v(\bx,t_{i})=\dot D_1(t_i)/\dot D_1(t_{i-1})v(\bx,t_{i-1}) $, i.e. we evolve the velocity field with its linear growth factor. Here $D_1$ is the linear density growth factor and the continuity equation implies $\theta =-\dot\de$. The logarithmic growth rate is defined as 
\be\label{e:deff}
f = \HH^{-1}\dot D_1/D_1 \quad \mbox{ hence } \theta = -f\HH\de 
\ee
within linear perturbation theory.
We call this method the '{\it rescaled} method'.
\item The fourth method consists of a Gaussian smoothing for both the momentum density and the density field. The velocity field is then estimated as the ratio of the two smoothed fields.
This method requires the introduction of a typical scale which sets the size of the smoothing.
We also note that a Gaussian smoothing of the fields in real space is not practical, as this would completely destroy the parallelizability of the code, which relies on the fact that each processor needs to know the particle positions only in a small region of the grid (and its \emph{halo}~\cite{Adamek:2016zes}). We therefore smooth in Fourier space, where the real space convolution becomes a simple multiplication. This requires two additional Fourier transforms of $T^0_0$ and $T_0^i$, but only at the time steps at which we want to determine the velocity power spectrum.
\end{enumerate} 
In section \ref{s:res} we shall show that the first method somewhat overestimates the vorticity power spectrum, while the three other methods agree to a good approximation, provided the smoothing scale is chosen sufficiently small, i.e. of the order of $1/10$ of a grid spacing or less.

We also compute the trace of the velocity dispersion tensor. The velocity dispersion $\sigma^2$ is computed similarly to the velocity field, i.e. we define
\be
\sigma^2 = -\frac{T^i_i}{T^0_0} - v^i v_i, \label{sig2}
\ee
where $T^i_i$, $T^0_0$ and $v^i$ are computed on the vertices of our grid cells with the CIC method \cite{Hockney:1988:CSU:62815}. 
The definition of $\sigma^2$, similarly to the velocity field, requires a prescription for estimating the squared velocity field $v^2 \equiv T^i_i/T^0_0$ in the voids.
We adopt for this computation the \emph{rescaled} method, which seems most physical, in particular we will rescale the squared velocity field with the square of the rescaling factor that we adopt for the velocity 
\be
v^2(\bx,t_{i})=[\dot D_1(t_i)/\dot D_1(t_{i-1})]^2 v^2(\bx,t_{i-1}). 
\ee 

In the next section we describe the decomposition of the velocity field in a divergence and a rotational part and the computation of the divergence and vorticity spectra.

\subsection{Decomposition of the velocity field}
\label{s:method2}

Since the N-body particles in \emph{gevolution} are projected onto a regular Cartesian grid, the Helmholtz decomposition of the velocity field is easily performed in Fourier space.

In Fourier space, the divergence $\theta$ and the vorticity $\boldsymbol{\om}$ are 
\bea
\theta(\bk) &=& k^2v(k),\\
\bom(\bk) &=& i\bk\wedge\bv_{R}(\bk).
\eea 
As usual, we denote a function and its Fourier transform by the same symbol. The Fourier transform is only indicated by the argument $\bk$.
In the code, the velocity field is decomposed into its divergence and rotational part
\be
\theta =   i (k_j v^j), \qquad \qquad
v^i_{R} =  P^{ij} v_j, 
\ee
where $i, j$ are index running over the Fourier space coordinates and $P^{ij}$ is the transverse projector,
$P^{ij}= \delta^{ij} -k^i k^j/k^2$.

In the numerical code, we have to replace the vector $\bk$ by the correspondent discrete wave vector,
\be
\bk_{{u},{v}, {w}} \equiv \frac{1}{\De_\text{grid}} \Biggl(\sin{ \frac{2 \pi {u}}{N}}, \sin{ \frac{2 \pi {v}}{N}}, \sin{ \frac{2 \pi {w}}{N}}\Biggr),
\ee
where $\De_\text{grid}$ is the grid spacing, $N$ is the number of grid points in each direction and 
$u,v,w$ are integers running from $1$ to $N$.

In a spatially homogeneous Universe the power spectra of the velocity, divergence and vorticity are of the form

\begin{align}
\mean{v^i(\mathbf{k}) v^{*\,j}(\mathbf{k'}) }&= \delta(\mathbf{k}-\mathbf{k'})(2\pi)^3 \left(\delta^{ij}P_1(k) + \frac{k^i k^j}{k^2}P_2(k)\right)\, ,  \label{e:Pvtot}\\
\mean{v^i_\text{R} (\mathbf{k}) v^{*\,j}_\text{R} (\mathbf{k'}) }&= \delta(\mathbf{k}-\mathbf{k'})(2\pi)^3\frac{1}{2}  \left(\delta^{ij} -  \frac{k^i k^j}{k^2}\right)P_R(k)\, ,  \label{e:Pvrot}\\
\mean{v^i_\text{G} (\mathbf{k}) v^{*\,j}_\text{G} (\mathbf{k'}) }&= \delta(\mathbf{k}-\mathbf{k'}) (2\pi)^3  \frac{k^i k^j}{k^2}P_G(k)\,,  \label{e:PG}\\
\mean{v^i(\mathbf{k}) v_i^{*}(\mathbf{k'}) }&= \delta(\mathbf{k}-\mathbf{k'})(2\pi)^3 P_{v}(k)\, ,\\
\mean{\theta (\mathbf{k}) \theta^{*} (\mathbf{k'}) }&= \delta(\mathbf{k}-\mathbf{k'}) (2\pi)^3P_\theta(k)\, ,\\
\mean{\om^i (\mathbf{k}) \om^{*\,j} (\mathbf{k'}) }&= \delta(\mathbf{k}-\mathbf{k'})(2\pi)^3 \frac{1}{2}  \left(\delta^{ij} -  \frac{k^i k^j}{k^2}\right)P_\om(k)\,. \label{e:Pom}
\end{align}
The Dirac delta is a consequence of statistical homogeneity.
The expression $\delta^{ij}P_1(k) + k^i k^j/k^2P_2(k)$ of \eqref{e:Pvtot} is the most general isotropic vector power spectrum, and the prefactors of $P_R$ in \eqref{e:Pvrot} and of $P_\om$ in \eqref{e:Pom} ensure $k_iv_R^i=0$ and $k_i\om^i=0$, respectively.
The prefactor of $P_G$ in \eqref{e:PG} makes use of the fact that $\bv_G$ is a gradient field.
Of course these spectra are not independent but enjoy the following relations which are easily verified:
\bea
P_R &=& 2P_1,\\
P_G &=& P_2+P_1, \\
P_v &=& 3P_1 + P_2 = P_R +P_G, \\
P_\theta &=& k^2P_G,\\
P_\om &=& k^2P_R, \label{PkVort} \\
k^2P_v &=& P_\theta+P_\om \,.
\eea

In the code, we implement the power spectrum of the rotational velocity $P_R(k)$.
However, in the next section we will show our results in terms of the vorticity power spectrum $P_\om$, which is computed from \eqref{PkVort}.

\subsection{Method comparison}

We tested the four methods described in section \ref{s:method2} by running a simulation for each method with the same parameters:
box size $L = 256\, \text{Mpc}/h$, which corresponds to a volume $V = L^3$; number of gridpoints, $N_\mathrm{grid}^3 = 512^3$; number of particles $N_\mathrm{part}^3 = 1024^3$.
We will refer to these simulation settings as \emph{low-resolution}.
The initial conditions of these simulations have been generated from the same seed.

\begin{figure}[h!]
\begin{subfigure}[b]{0.4\textwidth}
\includegraphics[width=\textwidth]{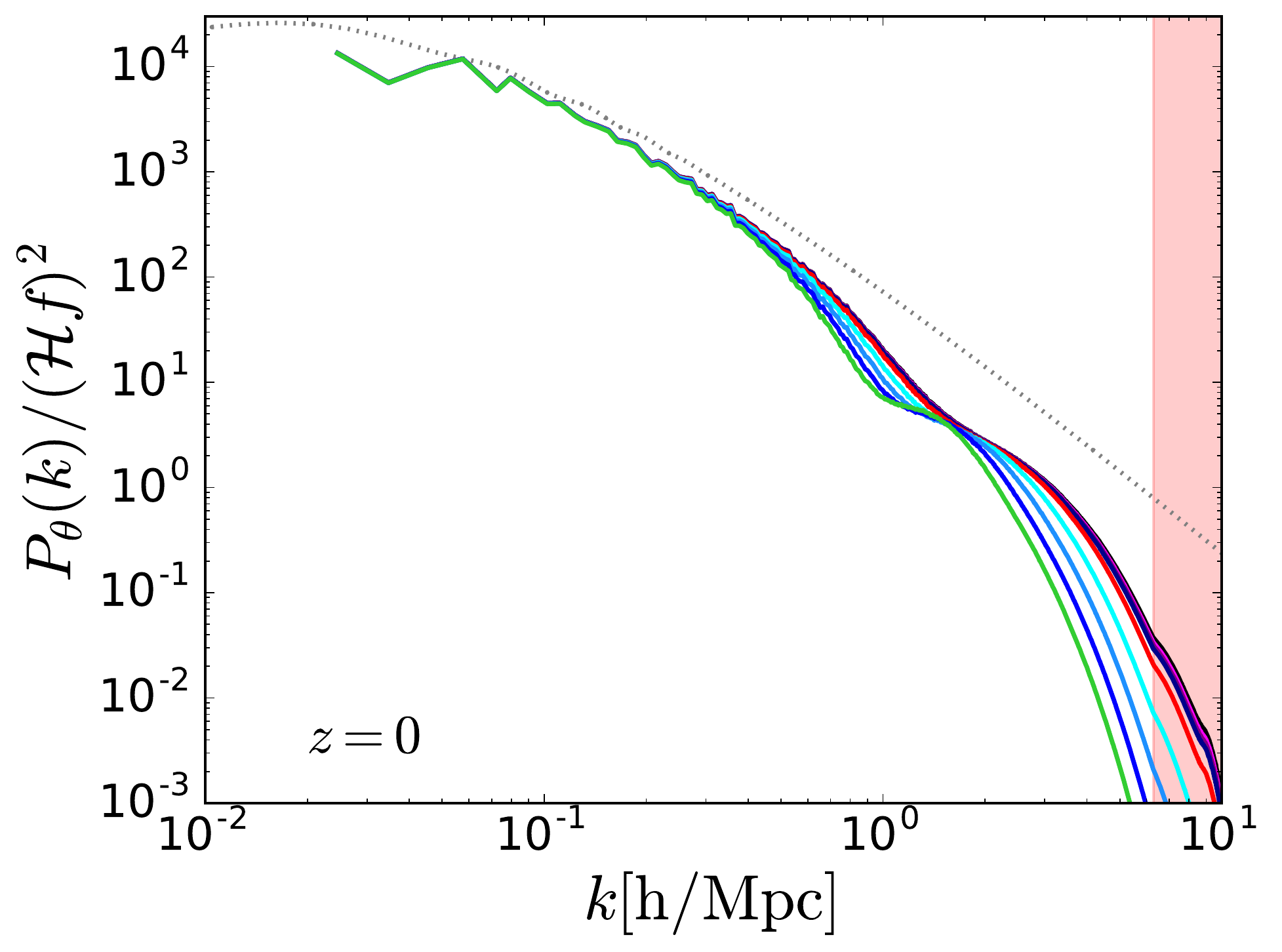}
\label{smooth:th0}
\end{subfigure}
\hspace{-0.2cm}
\begin{subfigure}[b]{0.4\textwidth}
\includegraphics[width=\textwidth]{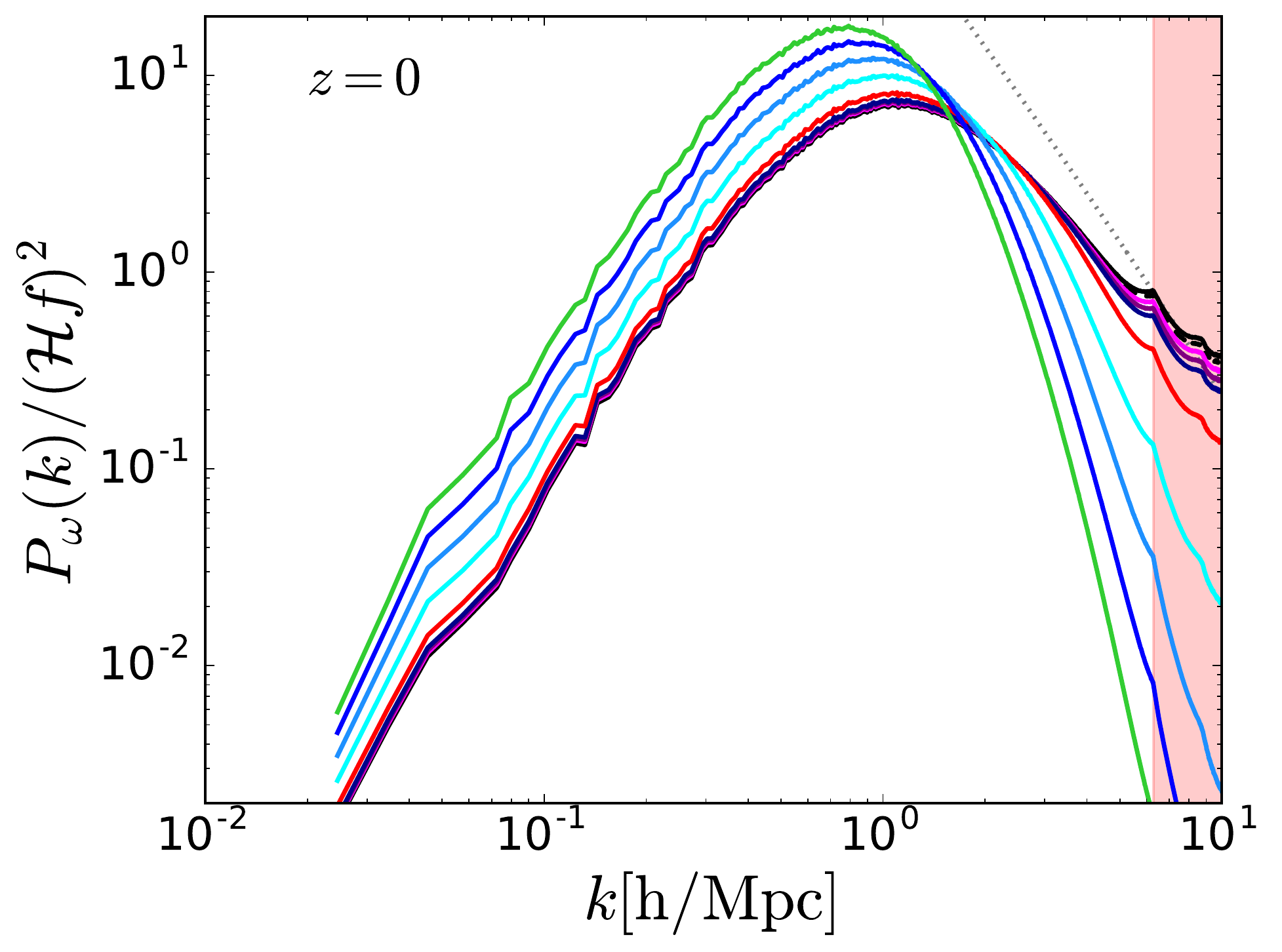}
\label{smooth:w0}
\end{subfigure}
\hspace{-0.2cm}
\begin{subfigure}[b]{0.2\textwidth}
\raisebox{0.1\height}{\includegraphics[width=\textwidth]{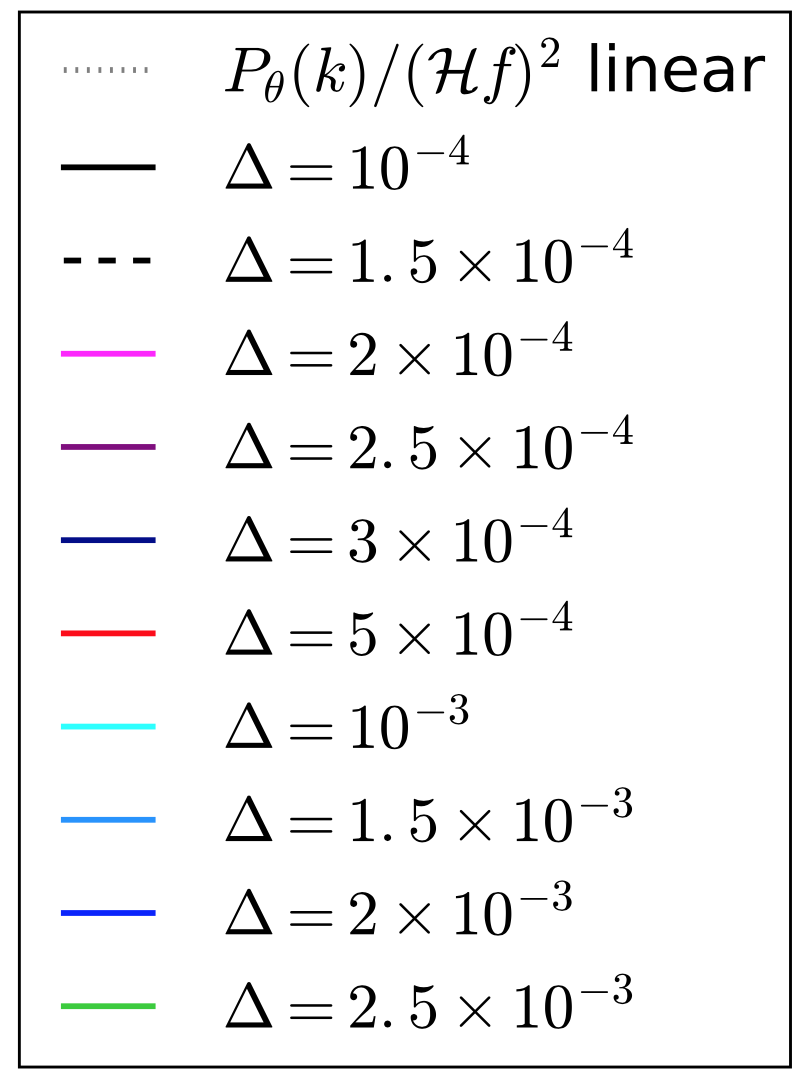}}
\label{smooth:legend.png}
\end{subfigure}
\\
\begin{subfigure}[b]{0.4\textwidth}
\includegraphics[width=\textwidth]{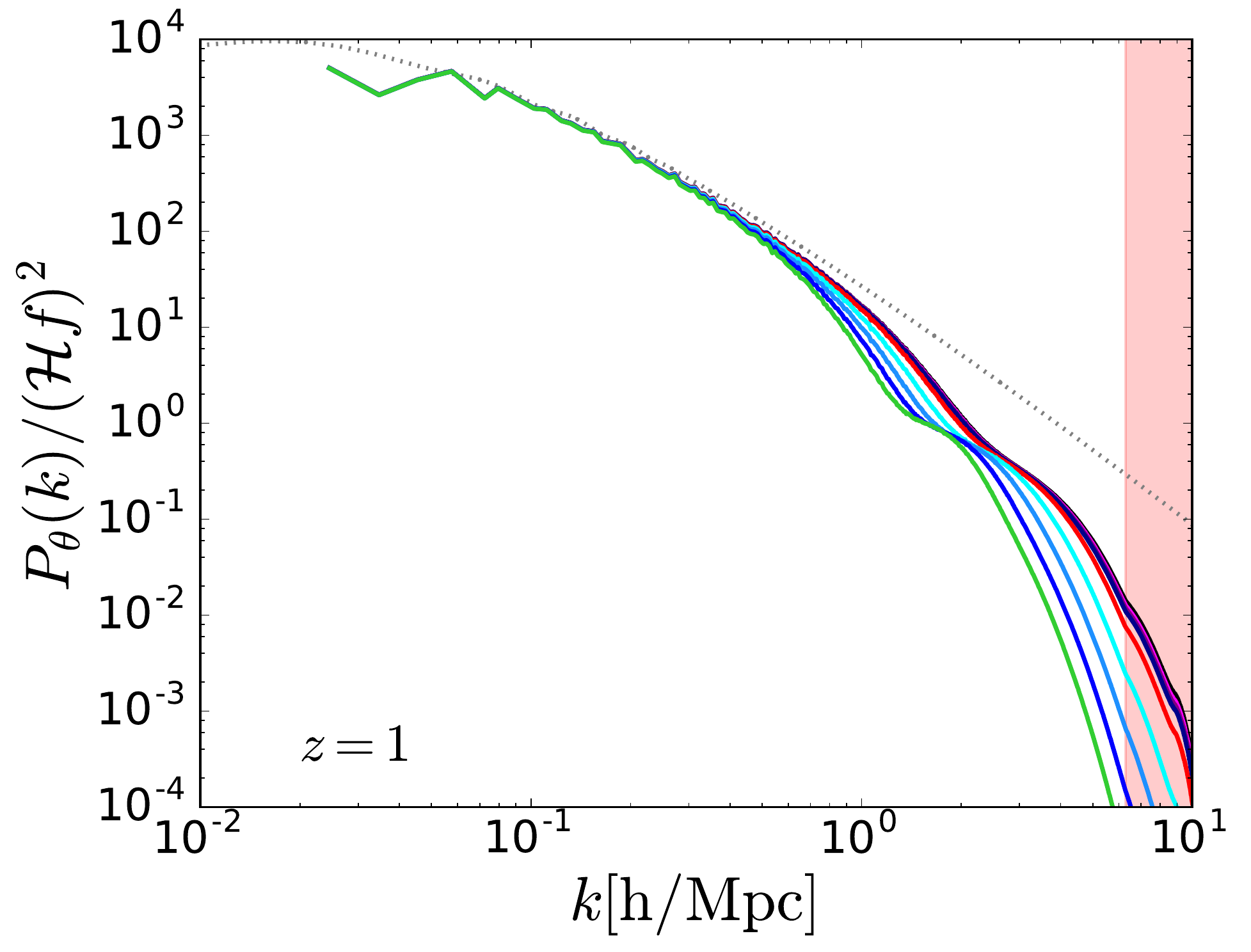}
\caption{Power spectrum - $\theta$}
\label{smooth:th1}
\end{subfigure}
\hspace{-0.2cm}
\begin{subfigure}[b]{0.4\textwidth}
\includegraphics[width=\textwidth]{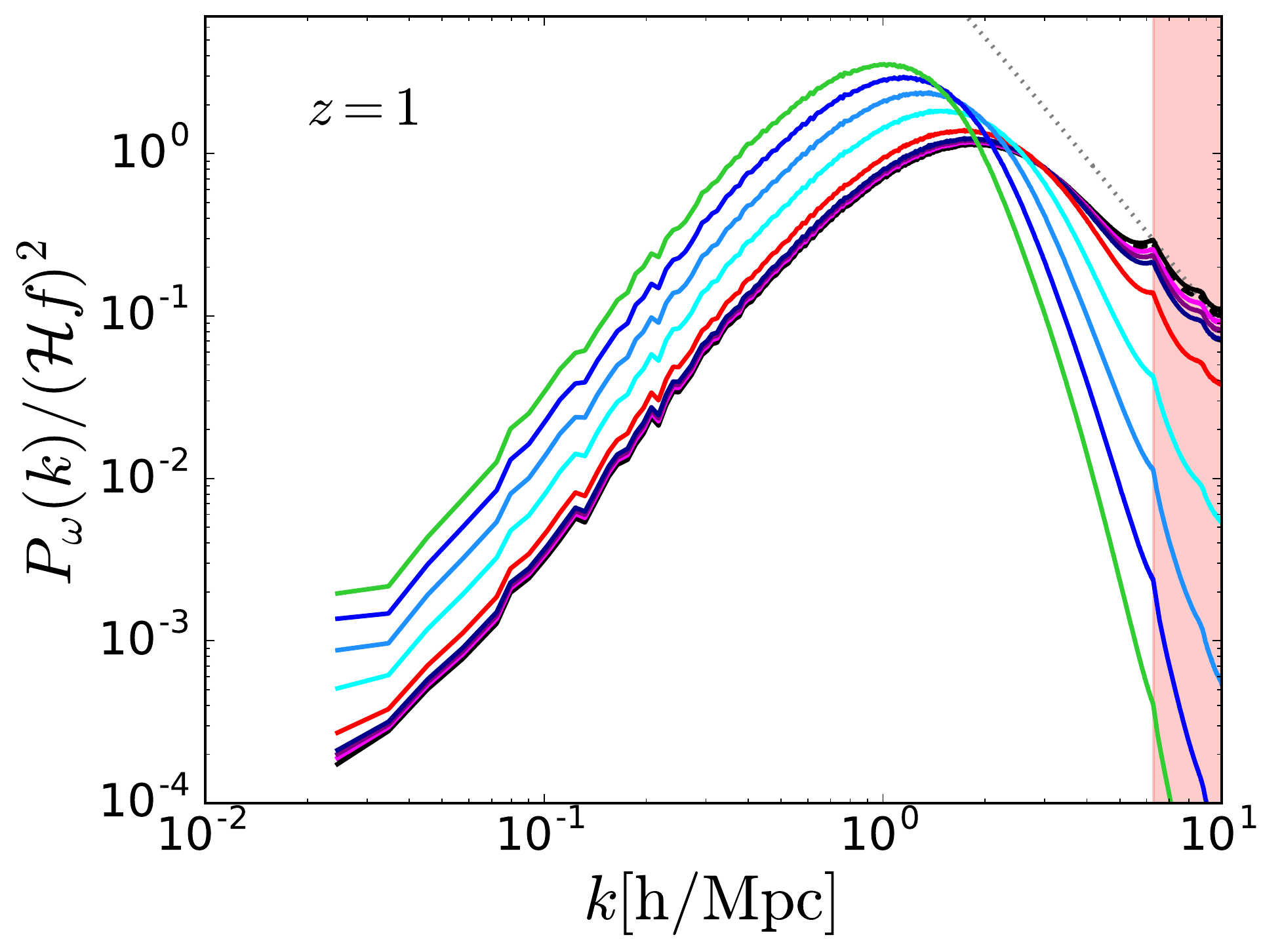}
\caption{Power spectrum - $\omega$}
\label{smooth:w1}
\end{subfigure}
\caption{Power spectrum of the divergence $\theta$ (left panel) and the vorticity $\omega$ (right panel) of the velocity field at $z = 1$ (top panels) and $z = 0$ (bottom panels). 
The power spectra are computed by applying a Gaussian smoothing, with standard deviation $\De$ (in units of the size of the box), to both the momentum density and density field.
Different colors refer to different smoothing scales.
A smoothing of the size of a grid cell corresponds to $\De=1/512 \approx 2 \times 10^{-3}$.
The fields are normalized in such a way that, on linear scales, $P_\theta$ matches the matter power spectrum. The red shaded region represents modes beyond the Nyquist frequency.
}
\label{f:smooth}
\end{figure}

In figure \ref{f:smooth} we show the power spectra of the divergence (left panels) and the vorticity (right panels), computed from the smoothed velocity field.
The smoothing is performed with a Gaussian filter, therefore this method has an extra degree of freedom that we have to test, i.e. the standard deviation of the Gaussian which sets the smoothing scale, which we denote as $\De$.
We run 10 simulations for different values of the smoothing scale, starting from $\De \approx 2 \times 10^{-3}$, which is approximately the size of one grid cell, down to $\De \approx 10^{-4}$, i.e. a smoothing size of $1/20$ the resolution of this set of simulations.
Top and bottom panels refer to different redshift, $z=0$ and $z=1$ respectively. 
The divergence spectra do not depend on $\De$ on large scales, while we observe a substantial suppression of power for $k \gtrsim 0.3h/$Mpc when increasing the smoothing scale from $10^{-4}$ to $2.5\times 10^{-3}$.
On the other hand, the vorticity power spectrum depends on the size of the smoothing also on large scales.
In particular, there is a transfer of power from small to large scales, when we increase the value of $\De$.
This is due to the fact that a relatively large smoothing scale (i.e. the size of a grid cell) introduces a spurious vorticity in our computation, since the velocity field at a given grid point is receiving a non-negligible contribution from particles located in the other regions of the simulation domain.
Since the vorticity is a purely non-linear effect, and in our cold dark matter simulations it is generated by orbit crossing of the particles, the smoothing of small scales introduces additional velocity dispersion which sources vorticity, see~\cite{Cusin:2016zvu}.
However, this spurious effect is small for a sufficiently small smoothing scale, and from figure \ref{f:smooth} we infer that the vorticity power spectrum converges if we choose a smoothing scale of $1/10$ the size of a grid cell in our simulation.
For this value the spurious vorticity induced by the smoothing is negligible.
Therefore, from now on, we will refer to the method \textit{smooth} to be the smoothing with a Gaussian kernel and a standard deviation of $1/10$ the size of a grid cell.

\begin{figure}[h!]
\centering
\begin{subfigure}[b]{0.495\textwidth}
\includegraphics[width=\textwidth]{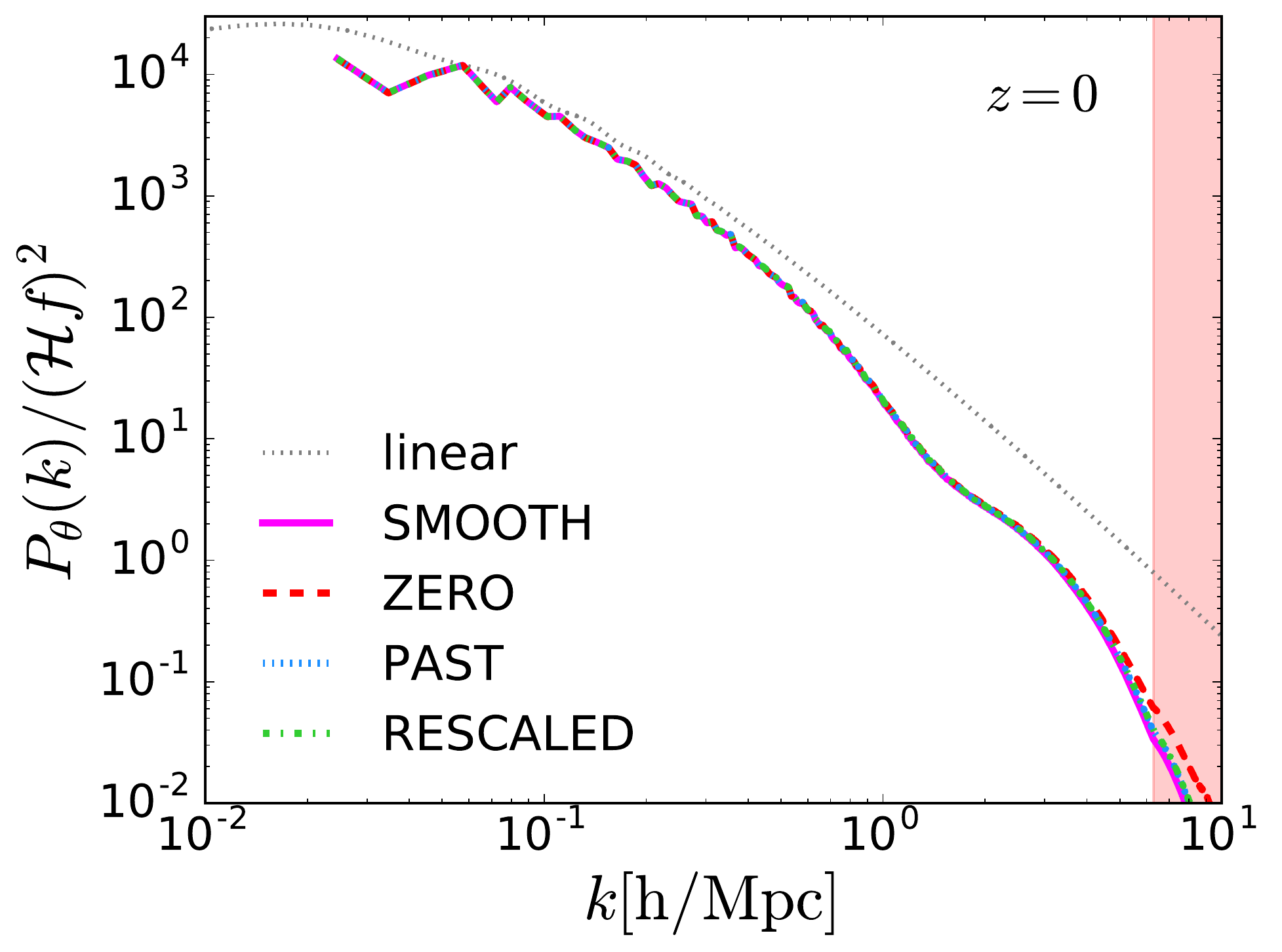}
\label{meth:th1}
\end{subfigure}
\begin{subfigure}[b]{0.495\textwidth}
\includegraphics[width=\textwidth]{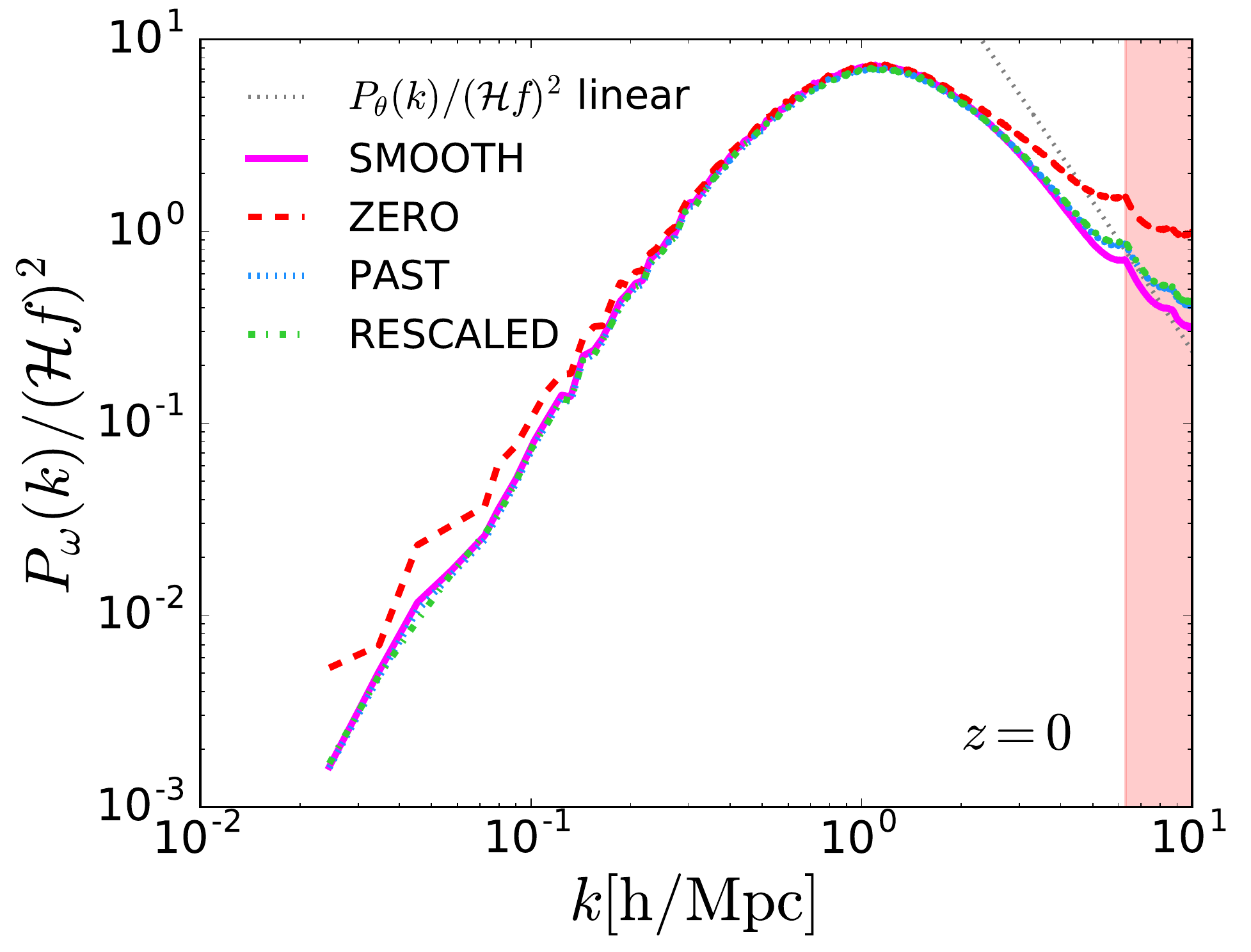}
\label{meth:w1}
\end{subfigure}
\\
\begin{subfigure}[b]{0.495\textwidth}
\includegraphics[width=\textwidth]{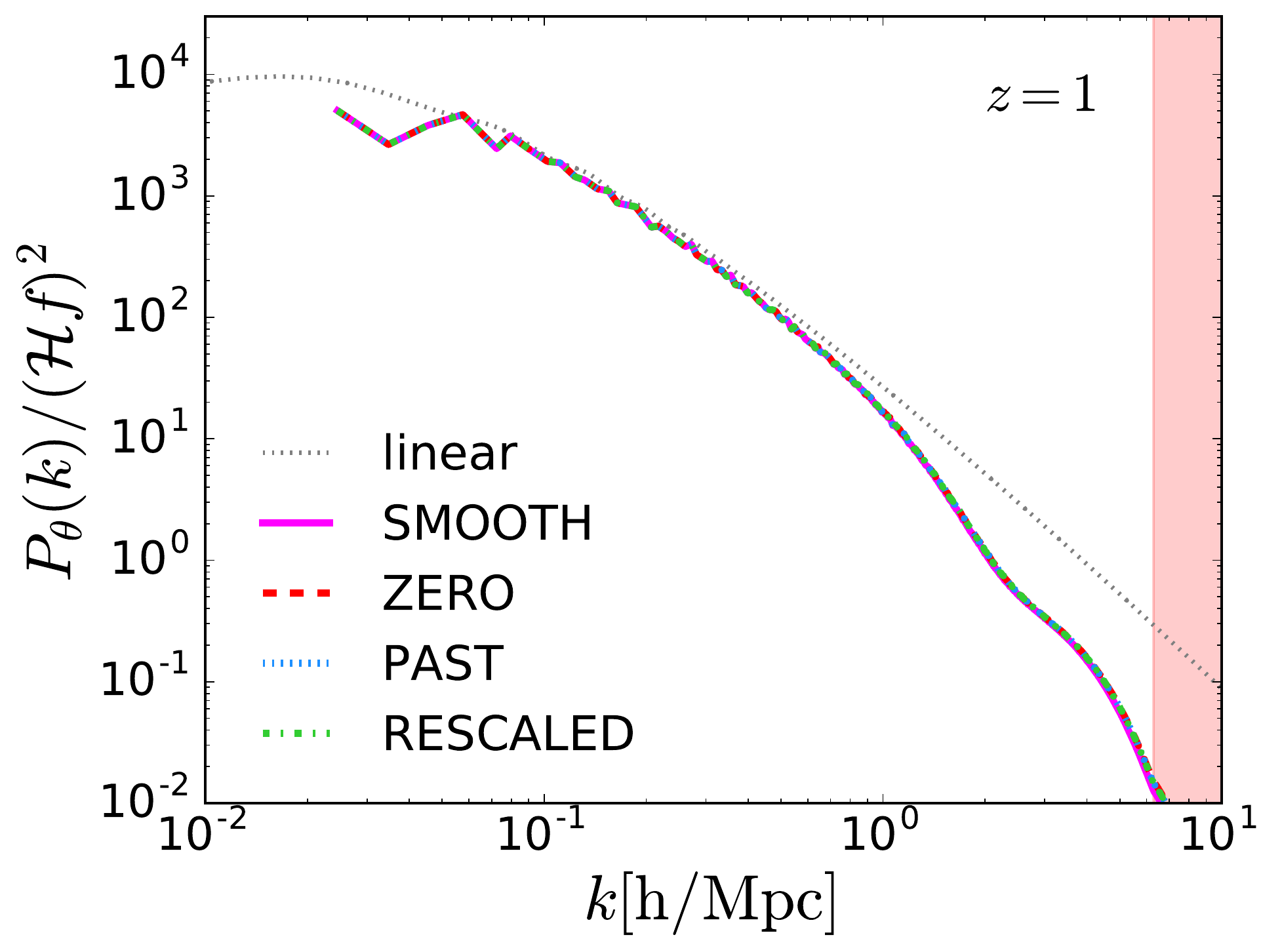}
\caption{Power spectrum - $\theta$}
\label{meth:th0}
\end{subfigure}
\begin{subfigure}[b]{0.495\textwidth}
\includegraphics[width=\textwidth]{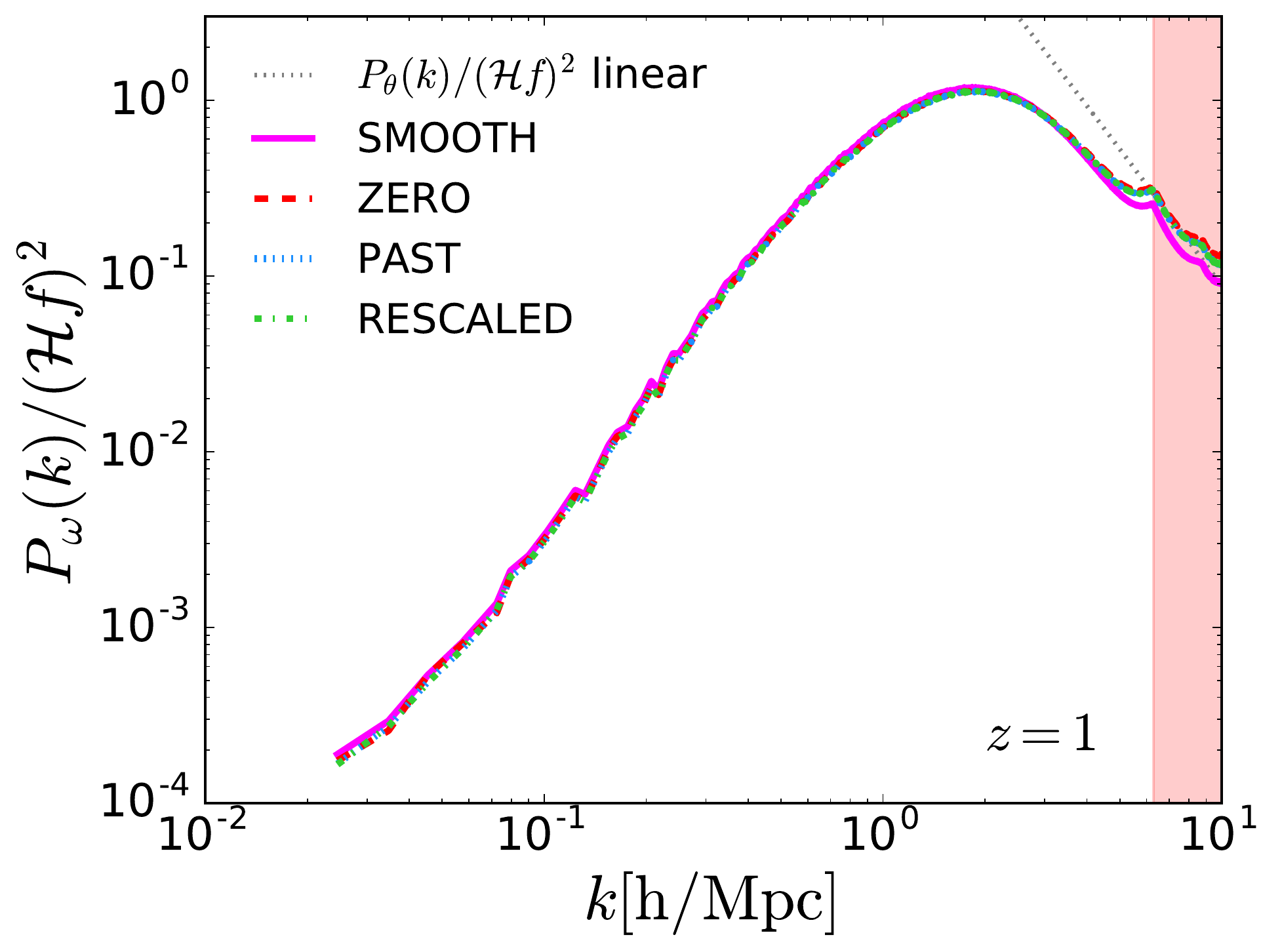}
\caption{Power spectrum - $\omega$}
\label{meth:w0}
\end{subfigure}
\caption{Power spectrum of the divergence $\theta$ (left panel) and the vorticity $\omega$ (right panel) field at $z = 0$ (top panels) and $z = 1$ (bottom panels). Different colors refer to the different methods for the velocity field computation. The method 'SMOOTH' refers to a Gaussian smoothing with width $\De = 2 \times 10^{-4}$ Mpc$/h$.
The spectra are normalized by dividing by $(f\HH)^2$ so that, on linear scales, $P_\theta$ matches the matter density power spectrum, see Eq.~\ref{e:deff}.
The results shown are for $L=256$ Mpc$/h$, $N_{\rm grid}=512$ and $N_\mathrm{part}=1024$.
The red shaded region represents the modes beyond the Nyquist frequency.}
\label{f:meth}
\end{figure}

In figure~\ref{f:meth} we compare the power spectra $P_\theta$ and $P_\om$ obtained with the four different methods to define the velocity field, at $z = 0$ (top panels) and $z=1$ (bottom panels).
At redshift $z=1$ all methods give consistent results. At $z = 0$ the differences between the four prescriptions are more prominent. However, all methods, 
apart from the one assigning $\bv=0$ to the empty points (red line), do agree up to the Nyquist frequency $k_\text{Nyq} = \pi N_\text{grid}/L$ (red shaded zone).

To summarize, we show that the \emph{past}, \emph{rescaled} and \emph{smooth} methods give the same results on all the scales relevant for our simulations. The \emph{rescaled} method is the physically best motivated prescription, and it has the advantage that it does not introduce any extra degrees of freedom that need to be tested for different parameter settings.
Therefore, unless stated otherwise, the simulations presented in the next section employ this prescription.

\section{Results}
\label{s:res}
In this section we collect the main results of this work.
The simulations discussed in this section are summarized in table \ref{Table-sims}.
All the simulations that will be discussed have the same physical box size of $256$ Mpc/$h$.
We tested for finite-volume effects by running a simulation with a larger box size and same spatial resolution $L/N_\text{grid}$ as the low-resolution simulations. The results are shown in appendix \ref{a:finbox}.
We found that the divergence and vorticity power spectra coincide in the two cases.
Therefore, the finite-volume effects for $L = 256$ Mpc/$h$ are negligible.
In section \ref{s:snap} we show the snapshots of our low-resolution simulations for the divergence, the vorticity and the velocity dispersion. In section \ref{s:largescale} we investigate the large-scale slope of the vorticity power spectrum, while in section \ref{s:converg} we show the result of the convergence test and we study the redshift dependence of the amplitude of the vorticity power spectrum. Finally, in section \ref{s:cross} we investigate the correlations of the velocity divergence with the density field and the 
velocity dispersion, and the correlations between the rotational part of the velocity and the vector perturbation of the metric $\mathbf{B}$.
In appendix \ref{a:dtfe} we compare our method for estimating the vorticity field with the Delaunay tessellation method (DTFE). We show that the DTFE method has a faster convergence than our method with respect to the grid resolution. However, we argue that the difference between the two methods does not impact on our final results which are obtained by higher resolution simulations than the ones used in the test.

\begin{table}[h!]
\centering
\begin{tabular}{|c|c|c|c|c|}
\hline
\multicolumn{1}{|c|}{}                                           & 
 \multicolumn{1}{c|}{$N_\text{grid}$}                   &
  \multicolumn{1}{c|}{$N_\text{part}$}                      &
  \multicolumn{1}{c|}{L\,  [Mpc/h]}                        &
  \multicolumn{1}{c|}{\#~ Realizations}                         \\
\hline
Low resolution             &    512     &  1024      &    256      &  16    \\
High resolution 1         &   1024    &   2048     &    256      &   1      \\
High resolution  2         &    2048   &   4096     &    256      &   1      \\
\hline
\end{tabular}
\caption{
Summary of the simulations discussed in sections~\ref{s:snap}, \ref{s:largescale}, \ref{s:converg}, \ref{s:cross}.
}
\label{Table-sims}
\end{table}

\subsection{Snapshots}
\label{s:snap}

\begin{figure}[h!]
\centering
\hspace{-0.3cm}
\begin{subfigure}[b]{0.33\textwidth}
\caption{$\theta$}
\includegraphics[width=\textwidth]{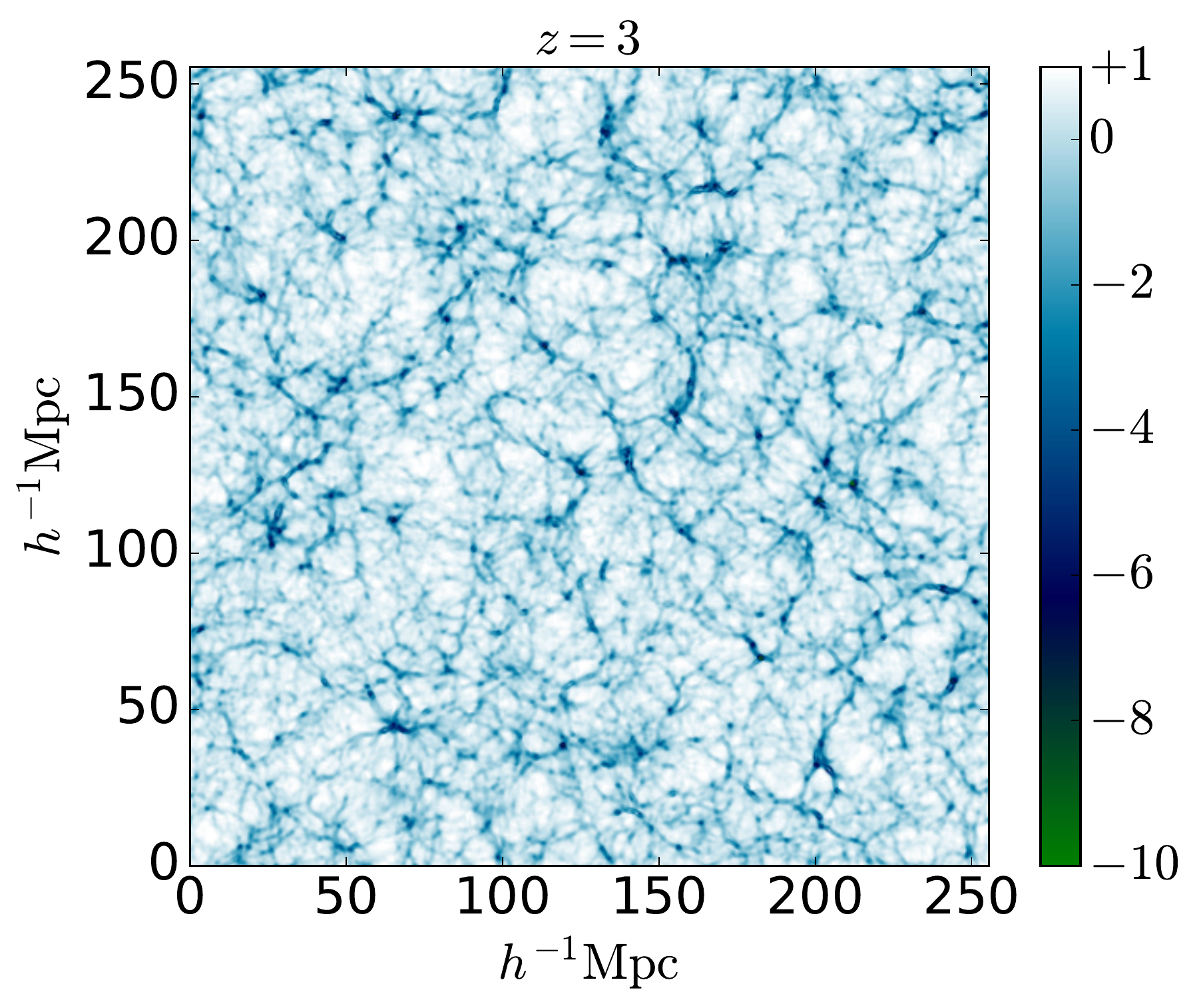}
\label{snap:theta}
\end{subfigure}
\hspace{-0.3cm}
\begin{subfigure}[b]{0.32\textwidth}
\caption{$|\om|$}
\includegraphics[width=\textwidth]{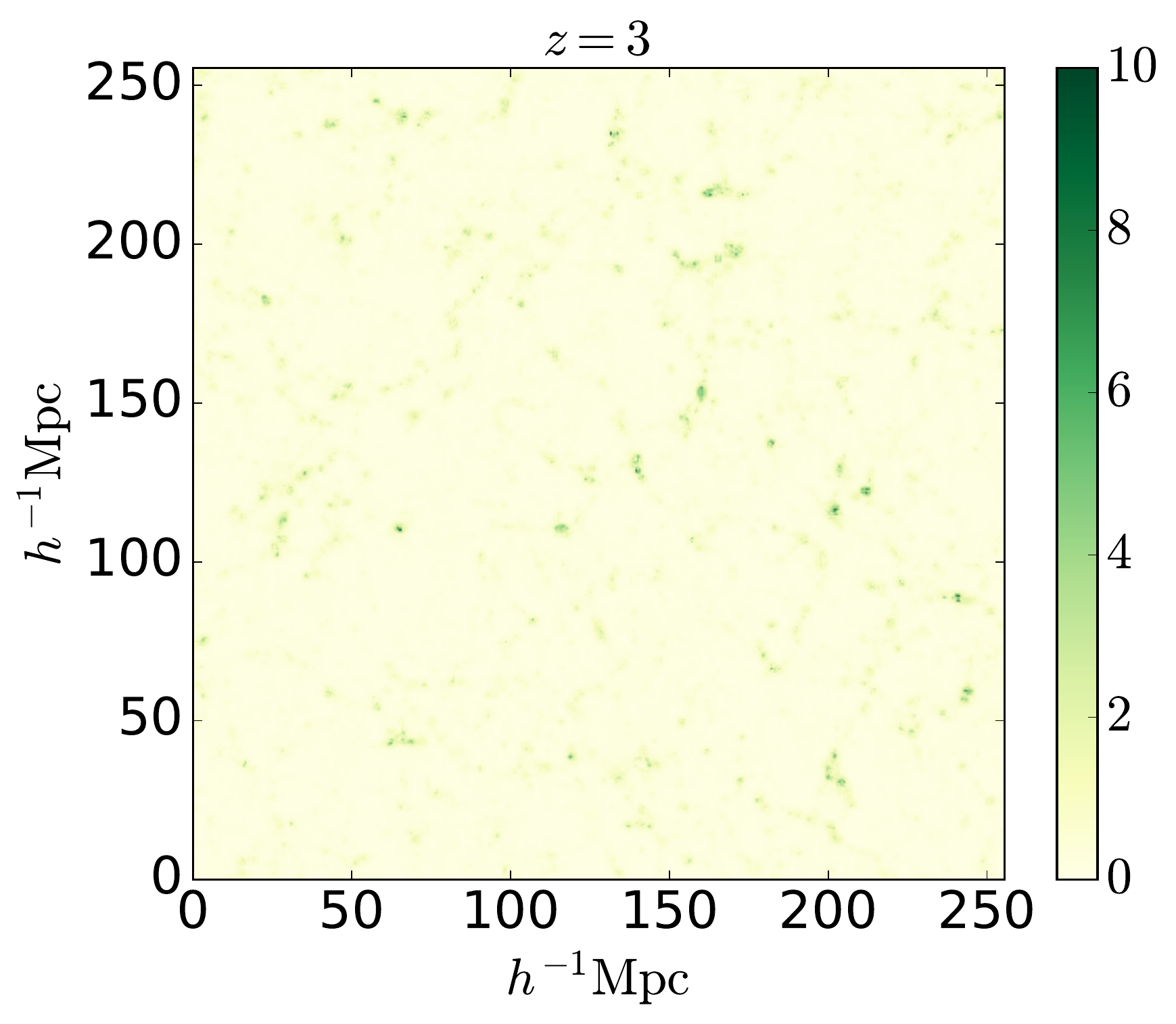}
\label{snap:omega}
\end{subfigure}
\hspace{-0.3cm}
\begin{subfigure}[b]{0.34\textwidth}
\caption{$\sigma^2$}
\includegraphics[width=\textwidth]{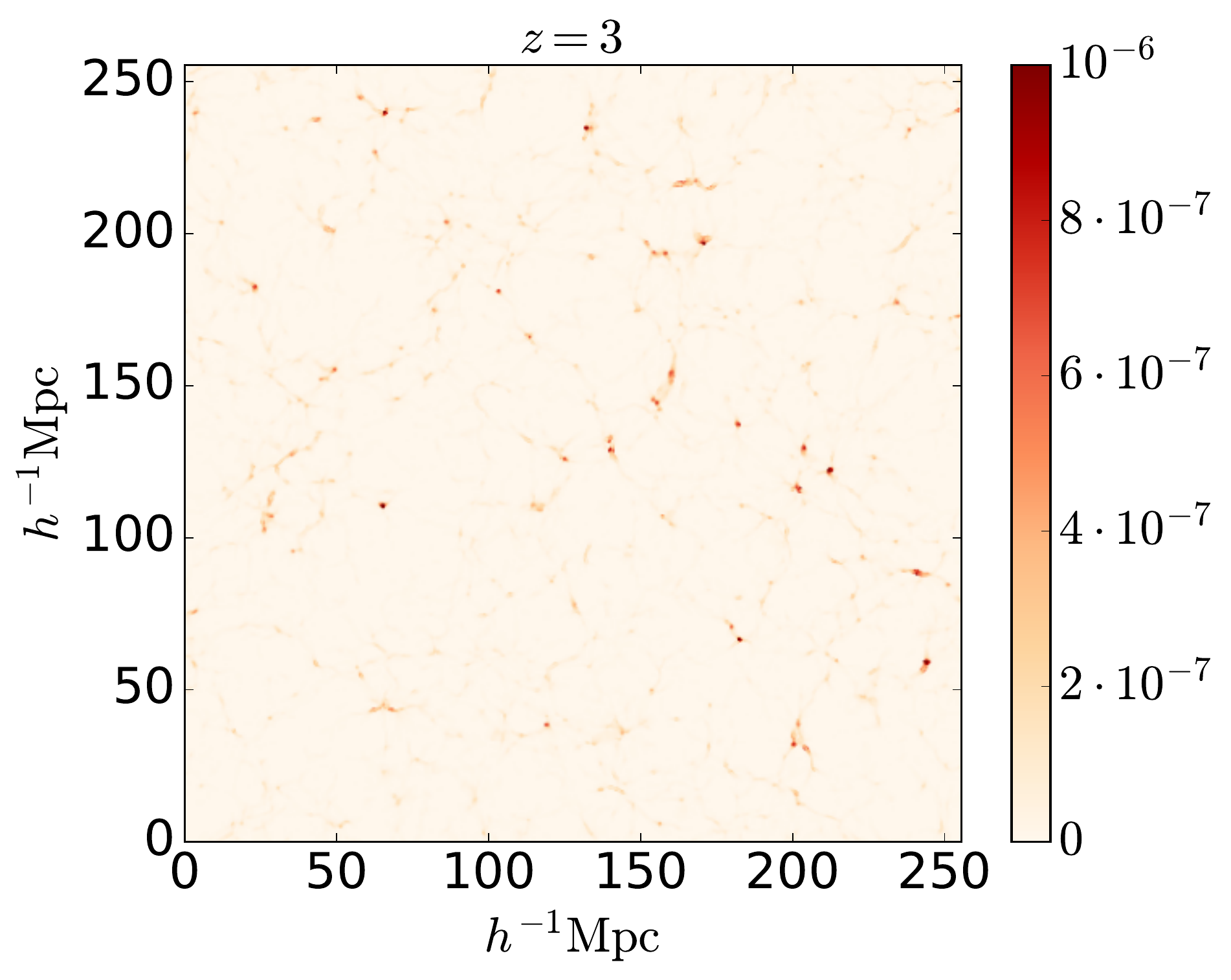}
\label{snap:sigma}
\end{subfigure}
\\
\hspace{-0.3cm}
\begin{subfigure}[b]{0.33\textwidth}
\includegraphics[width=\textwidth]{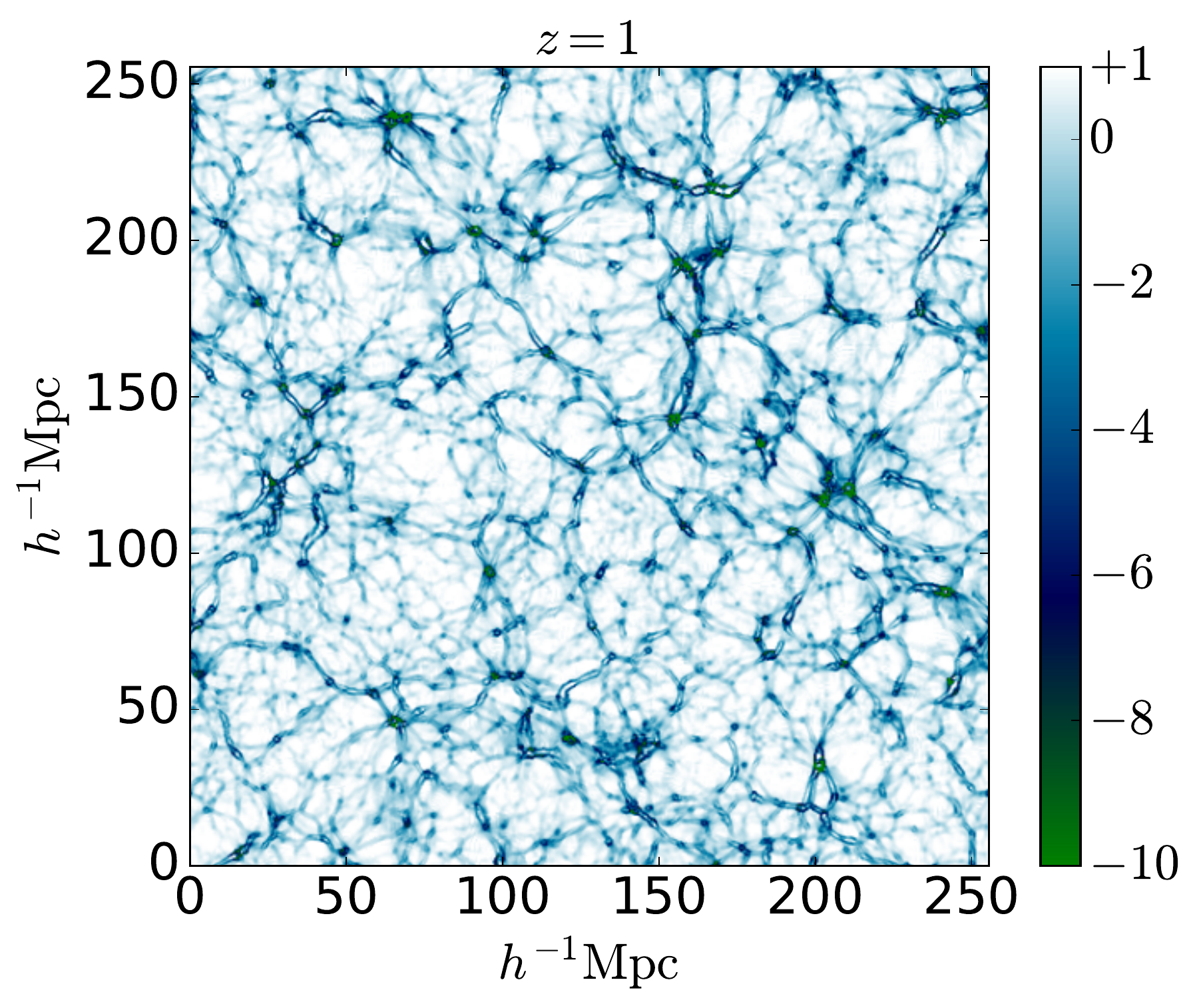}
\end{subfigure}
\hspace{-0.3cm}
\begin{subfigure}[b]{0.32\textwidth}
\includegraphics[width=\textwidth]{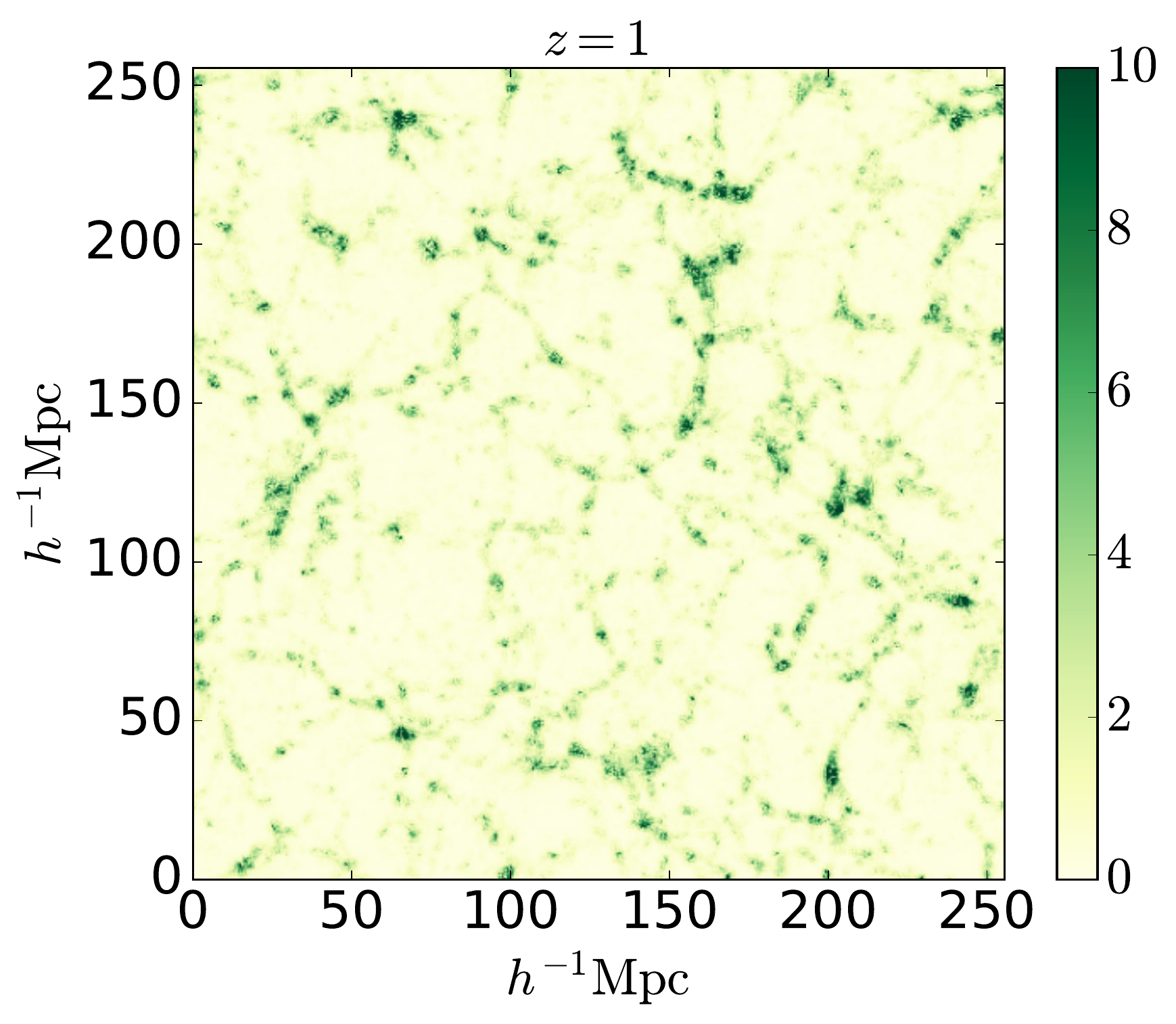}
\end{subfigure}
\hspace{-0.3cm}
\begin{subfigure}[b]{0.34\textwidth}
\includegraphics[width=\textwidth]{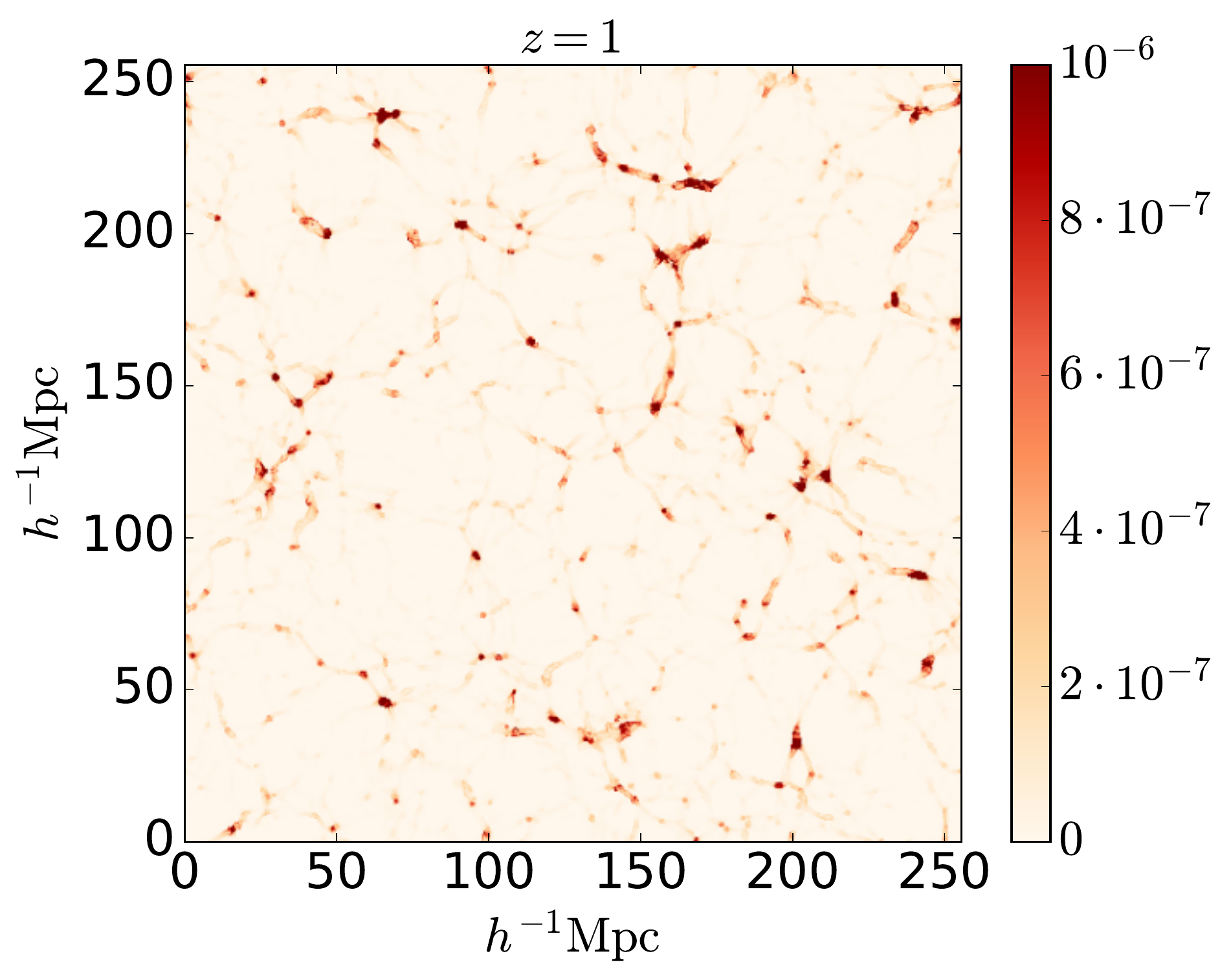}
\end{subfigure}
\\
\hspace{-0.3cm}
\begin{subfigure}[b]{0.33\textwidth}
\includegraphics[width=\textwidth]{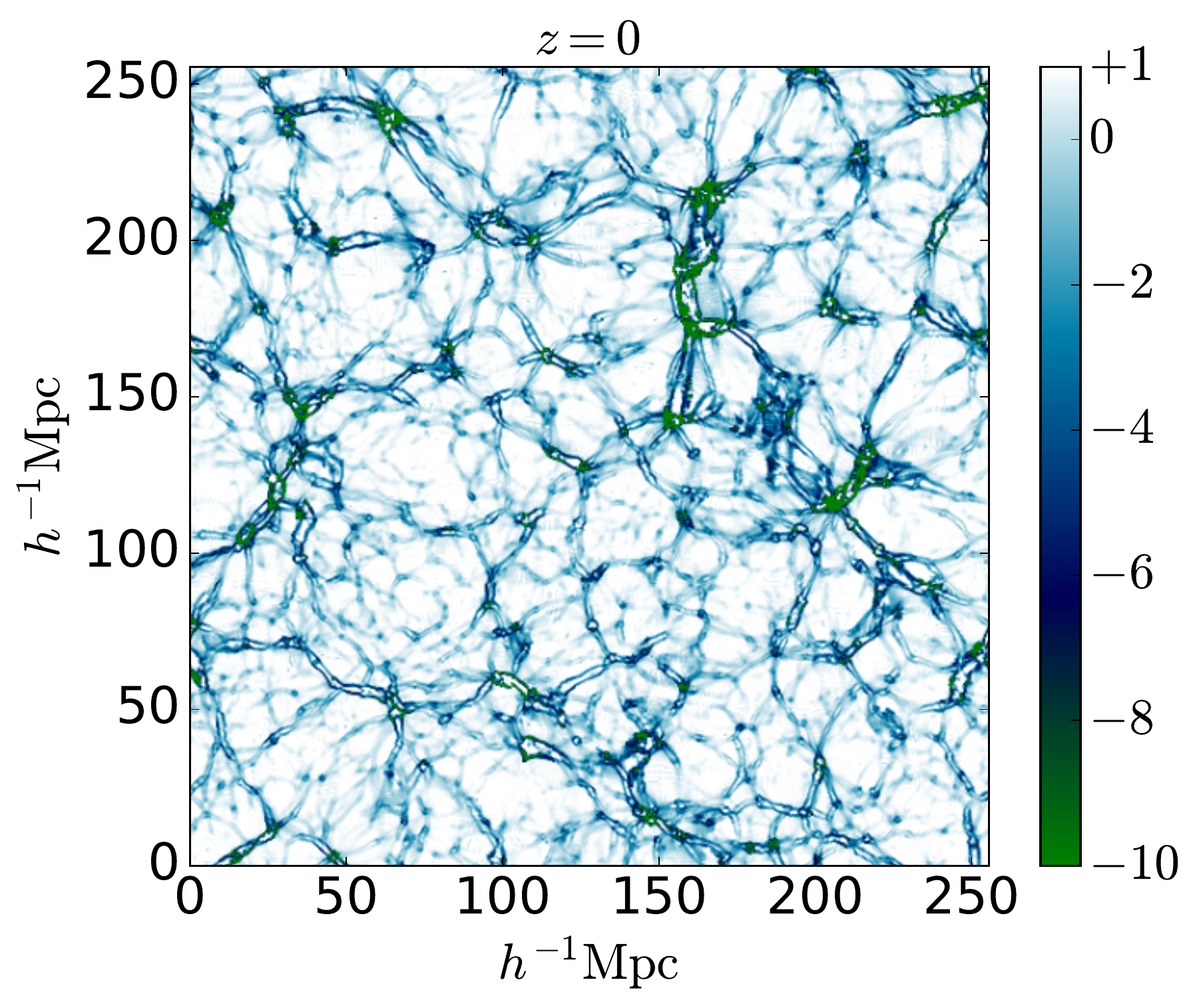}
\end{subfigure}
\hspace{-0.3cm}
\begin{subfigure}[b]{0.32\textwidth}
\includegraphics[width=\textwidth]{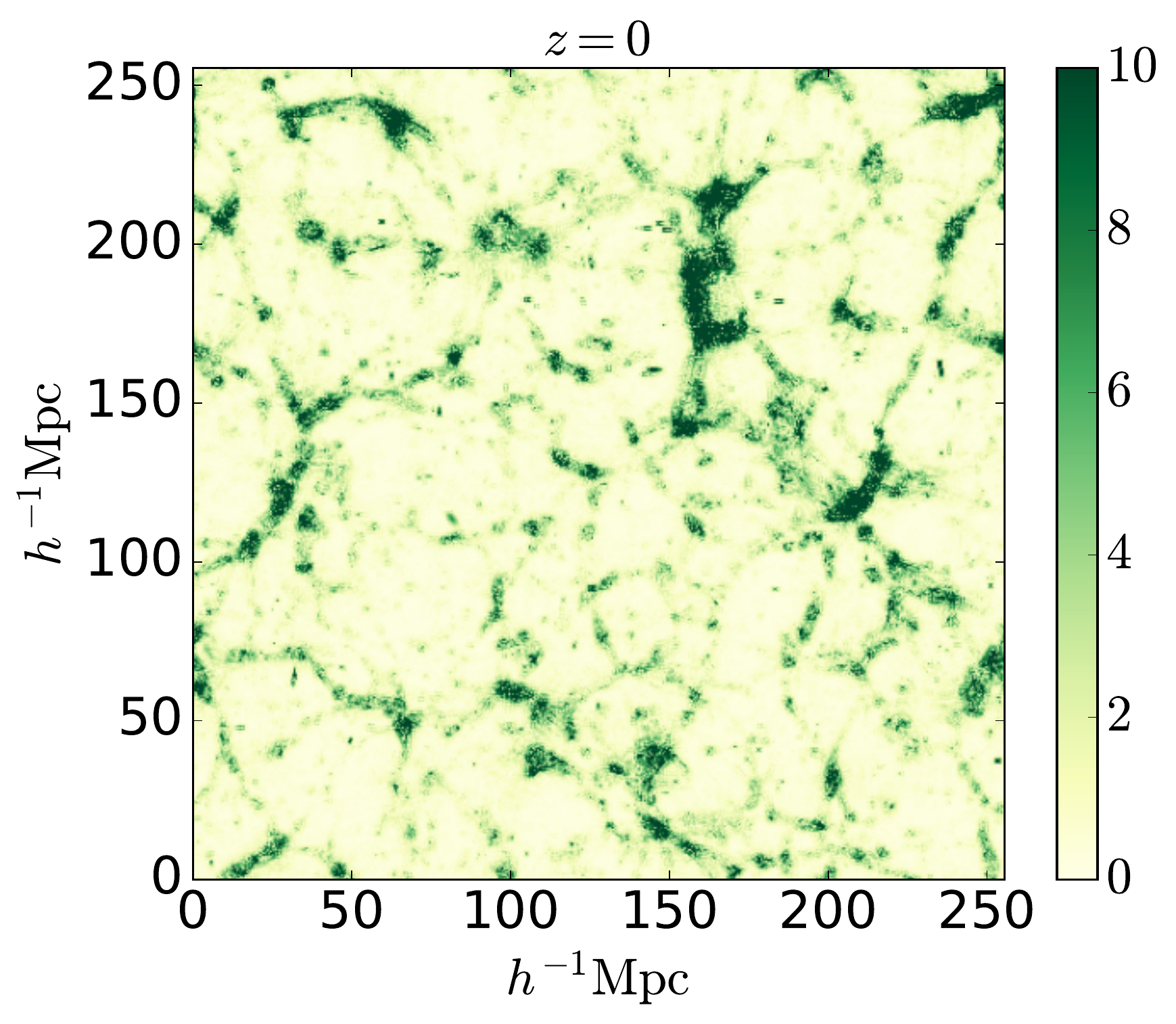}
\end{subfigure}
\hspace{-0.3cm}
\begin{subfigure}[b]{0.34\textwidth}
\includegraphics[width=\textwidth]{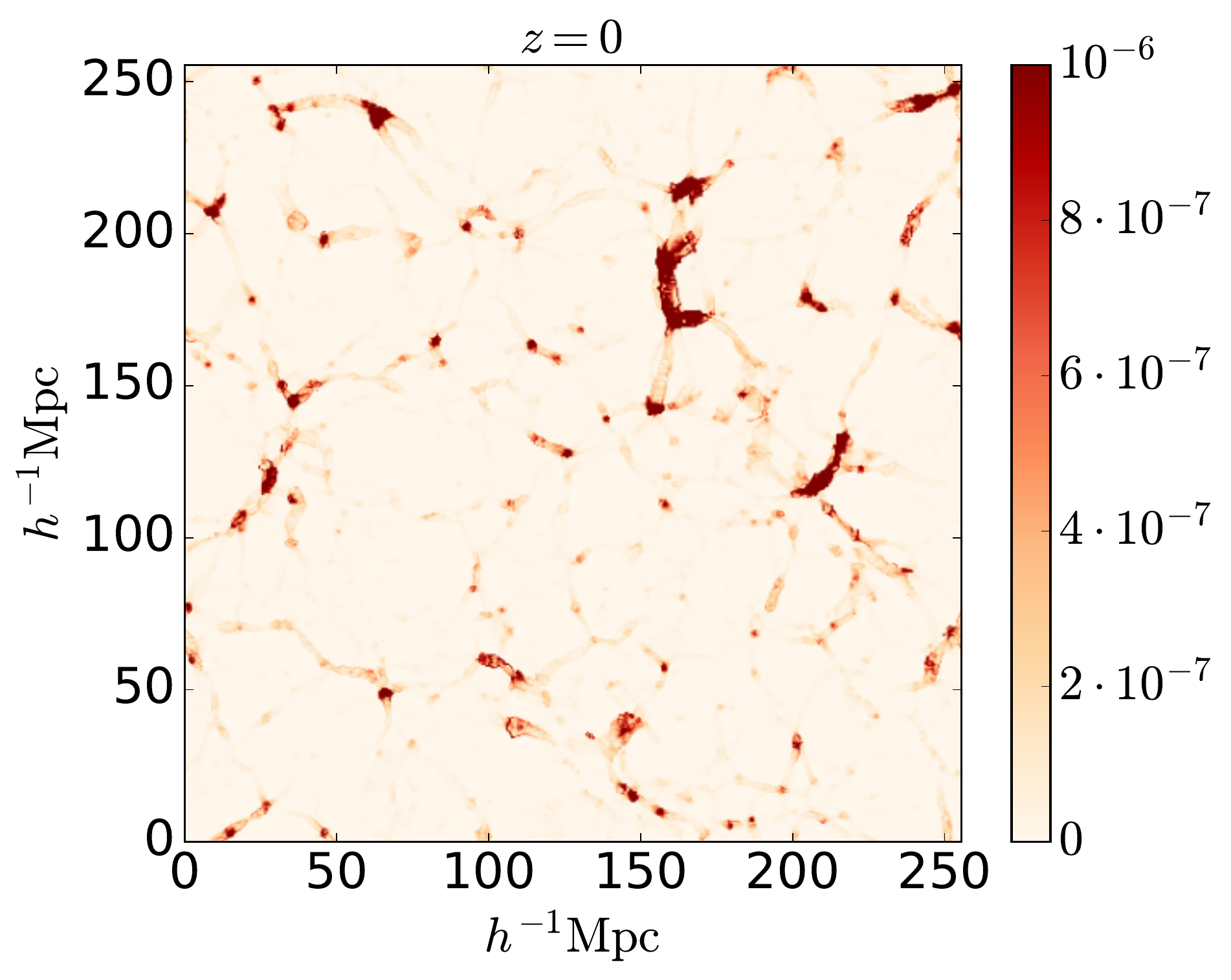}
\end{subfigure}
\caption{A slice through our simulation at $z=3$ (top panels), $z=1$ (middle panels), $z=0$ (bottom panels) showing the divergence field (left panels), the absolute value of the vorticity (middle panels)
and the velocity dispersion (right panels). }
\label{f:snaps}
\end{figure}

In figure~\ref{f:snaps} we show a slice through one of our low resolution simulations at redshift $z=3$ (top panels), $z = 1$ (center panels) and $z = 0$ (bottom panels).
We show the divergence, the vorticity and the velocity dispersion.
The velocity dispersion is a new result which has not been determined in previous simulations.
The divergence and the vorticity are normalized by the factor $(\HH f)$, therefore they are dimensionless, and the divergence field, at the linear level, satisfies $\theta/(\HH f) = -\delta$, where $\delta$ is the dark matter density fluctuation.

At first glance we can notice that the overall structures agree very well.
The divergence field is negative in the most over-dense regions, which translates into an inflow of particles, while it is positive in the voids, as expected. 
We also observe that both the vorticity and the velocity dispersion are peaked locally in the collapsing regions, where the orbit crossing of the particles occurs.
They have less large-scale power than $\theta$ which extends over the entire grid.
At $z=3$ both the vorticity and the velocity dispersion are very close to zero in the entire domain.
At $z = 3$ the mean velocity dispersion in the simulation is approximately $\bar{\sigma}^2 \approx 5\times 10^{-9}$, and it increases to $\bar{\sigma}^2 \approx 10^{-8}$ at $z = 1$ and to $\bar{\sigma}^2 \approx 3 \times 10^{-8}$ at $z = 0$. Hence it seems to grow somewhat faster than $1/(z+1)$, but not like $1/(z+1)^2$
as a background velocity dispersion would, as argued in~\cite{Cusin:2016zvu}.
\subsection{Large-scale behavior}
\label{s:largescale}

In order to study the large-scale behavior of the vorticity power spectrum, we run 16 realizations of the \emph{low-resolution} simulations described in table \ref{Table-sims}.
In fact, while the amplitude of the vorticity power spectrum is resolution dependent and needs a higher resolution simulation to be estimated, the spectral index is not affected by the resolution.
 
\begin{figure}[t]
\begin{center}
\includegraphics[width=0.6\textwidth]{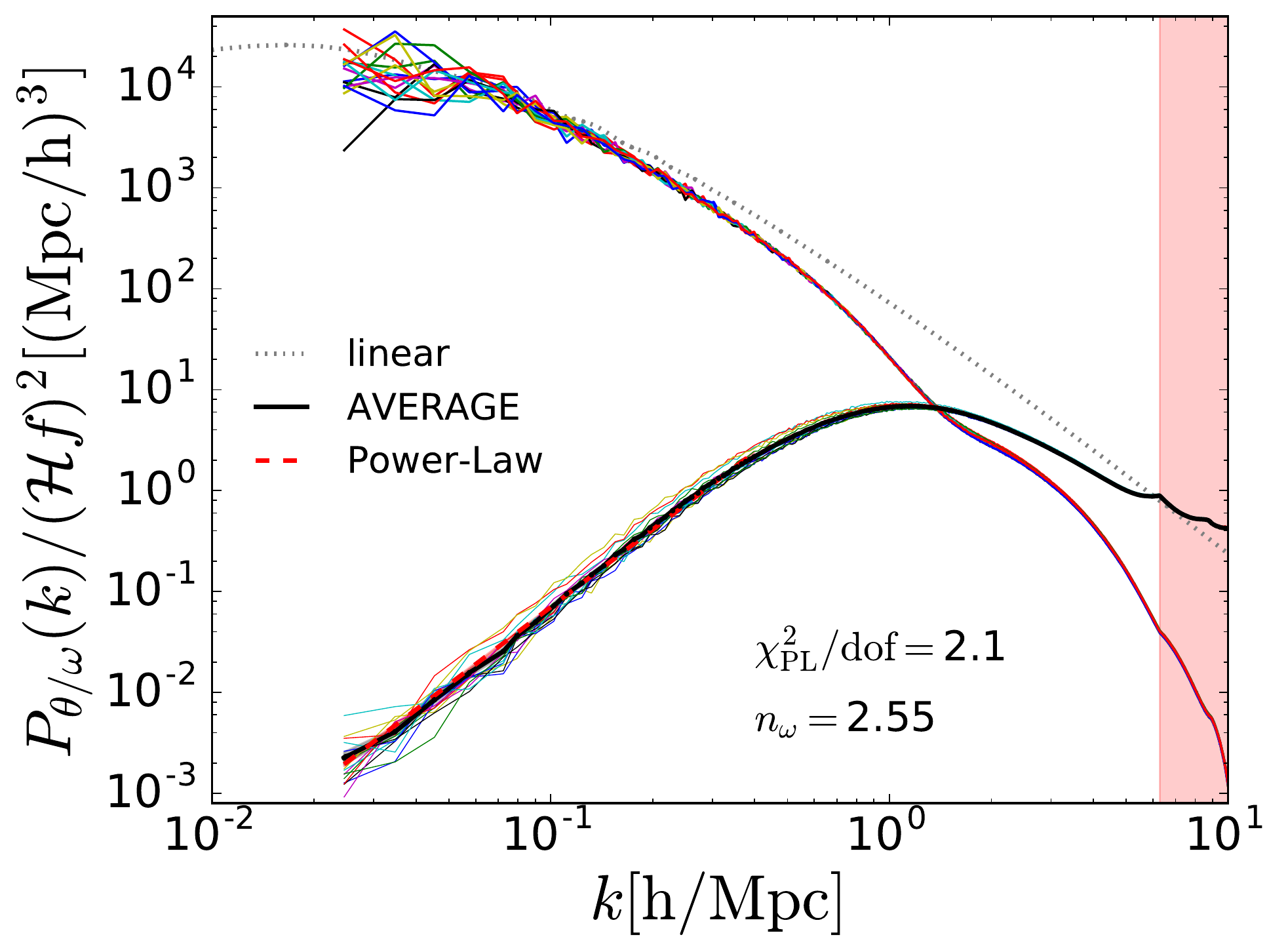}
\caption{
Divergence and vorticity power spectra for different realizations of the low-resolution simulation from table \ref{Table-sims}, at $z=0$. Different colors refer to different initial seeds.
The black continuous line represents the average over the 16 realizations, while the red dashed line represents the result of the power-law fit, with $k_\text{max} = 0.4\, h$/Mpc.
The $\chi^2$ divided by the degrees of freedom for the fit is approximately $2$.
}
\label{f:slope}
\end{center}
\end{figure}

In figure \ref{f:slope} we show the divergence and vorticity power spectra for these simulations.
Different colors denote different initial seeds.
The averaged power spectrum is represented by the black continuous line.
For $k \lesssim 0.4\, h$/Mpc the vorticity power spectrum is well modelled by a power-law with amplitude $A_\omega$ and spectral-index $n_\omega$,
\be
P_\omega(k) = A_\omega k^{n_\omega}. \label{pl}
\ee 

We have estimated the spectral index by fitting the model \eqref{pl} to the average power spectrum from our simulations.
The errors on the average power spectrum are computed as the standard deviation divided by the square root of the number of realizations. 

The red dashed line in figure \ref{f:slope} represents the result of our fit, and the best fit value for the spectral index is
\be
n_\omega = 2.55 \pm 0.02, \label{nw}
\ee
where the uncertainty quoted above is the statistical error. 
The best-fit value of the spectral index slightly depends on the $k-$range selected for the fit, with fluctuations of the order of $10\%$ in the range $k_\text{max} \in [0.3, 0.5]$ $h$/Mpc.

The values of the spectral index that we find is fairly consistent with previous results obtained from N-body simulations \cite{Pueblas:2008uv, Hahn:2014lca, Thomas:2015kua}, where
the simulations were performed with a different N-body code and the vorticity field was reconstructed with different techniques.
However, from a theoretical point of view, there is no unequivocal prediction for the vorticity power spectrum.

In the perturbative approach used in~\cite{Cusin:2016zvu}, the vorticity power spectrum, in the large-scale limit has a power-law behavior with spectral index $n_\omega = 4$.
However, in that analysis, vorticity is generated by the intrinsic dark matter velocity dispersion, while in our N-body approach, CDM has initially no velocity dispersion and it is the orbit crossing of particles that induces both, velocity dispersion and vorticity. 

In \cite{Scoccimarro:2000zr, Pueblas:2008uv} an expression for the vorticity power spectrum, generated by orbit-crossing, is computed analytically. Assuming that the vorticity that develops after shell-crossing is proportional to the curl of the mass-weighted velocity field, the vorticity power spectrum is expected to have a quadratic scale dependence in the low-$k$ limit, i.e. $n_\omega = 2$. 
This is also expected from simple causality: the initially vanishing vorticity can be correlated only in causally connected points. Therefore, the vorticity correlation function has compact support, hence its Fourier transform, the power spectrum is analytic for small $\bk$. The non-analytic pre-factor then requires that $P_\om \propto k^{2n}, ~ n \in {\mathbb N}$ (see e.g.~\cite{Durrer:2003ja} for more details). Without some special 'conservation law' we therefore expect $P_\om \propto k^2$ on large scales.
One might argue that the deviation from $n_\omega = 2$ comes from numerical contributions to the vorticity. But we have checked that our results are stable under increase of resolution (see Appendix B) and therefore are confident that the result at intermediate scales can be trusted. We rather suppose that we might miss some vorticity on the very largest simulated scales which would be induced from even larger scales not contained in the simulation. This would render the power spectrum too steep.

Results for the vorticity spectrum have also been obtained within the framework of the Effective Field Theory of Large Scale Structures (EFT of LSS) \cite{Carrasco:2012cv, Carrasco:2013mua}. The second of these references predicts $n_\omega = 2.8$ for $k \in [0.2, 0.6]$ $h$/Mpc, while the slope of the power spectrum becomes steeper on larger scales, i.e. $n_\omega = 3.6$ for $k \in [0.1, 0.2]$ $h$/Mpc.

It is also interesting to note that at $k\simeq 1$ $h/$Mpc where the vorticity power spectrum peaks, the divergence power spectrum has a dip and drops significantly below the linear perturbation theory value (dotted line in figure~\ref{f:slope}). Hence when non-linear structures form, part of the power in the divergence is converted into vorticity. We interpret this as the effect of angular momentum conservation which prevents further infall and forces the particles to rotate around a given structure.

\subsection{Resolution and redshift dependence}
\label{s:converg}
In this subsection, we describe some important tests we performed to ensure convergence, and consequently the reliability, of the results obtained from our simulations.
We remark that, since in \textit{gevolution} the fields are computed on a regular Cartesian grid, it is crucial to test the convergence of our results with respect to two quantities: the number of particles in each grid cell and the total number of particles (or total number of cells) in the box.
These two parameters are tested in sections \ref{s:converg1} and \ref{s:converg2}, respectively.
\subsubsection{On the choice of $N_\text{part}/N_\text{grid}$}
\label{s:converg1}
As $\emph{gevolution}$ is a code based on the particle-mesh (PM) scheme, the initial conditions are such that there are $R^3\equiv (N_\text{part}/N_\text{grid})^3$ particles in a single grid cell, and our results may depend on it, hence we need to determine what is a good value for $R$.
We ran several simulations for values $R=1,2,4$, with a fixed box size $L=256$ Mpc/$h$, and fixed $N_\text{grid}=512$.

\begin{figure}[h!]
\centering
\includegraphics[scale=0.5]{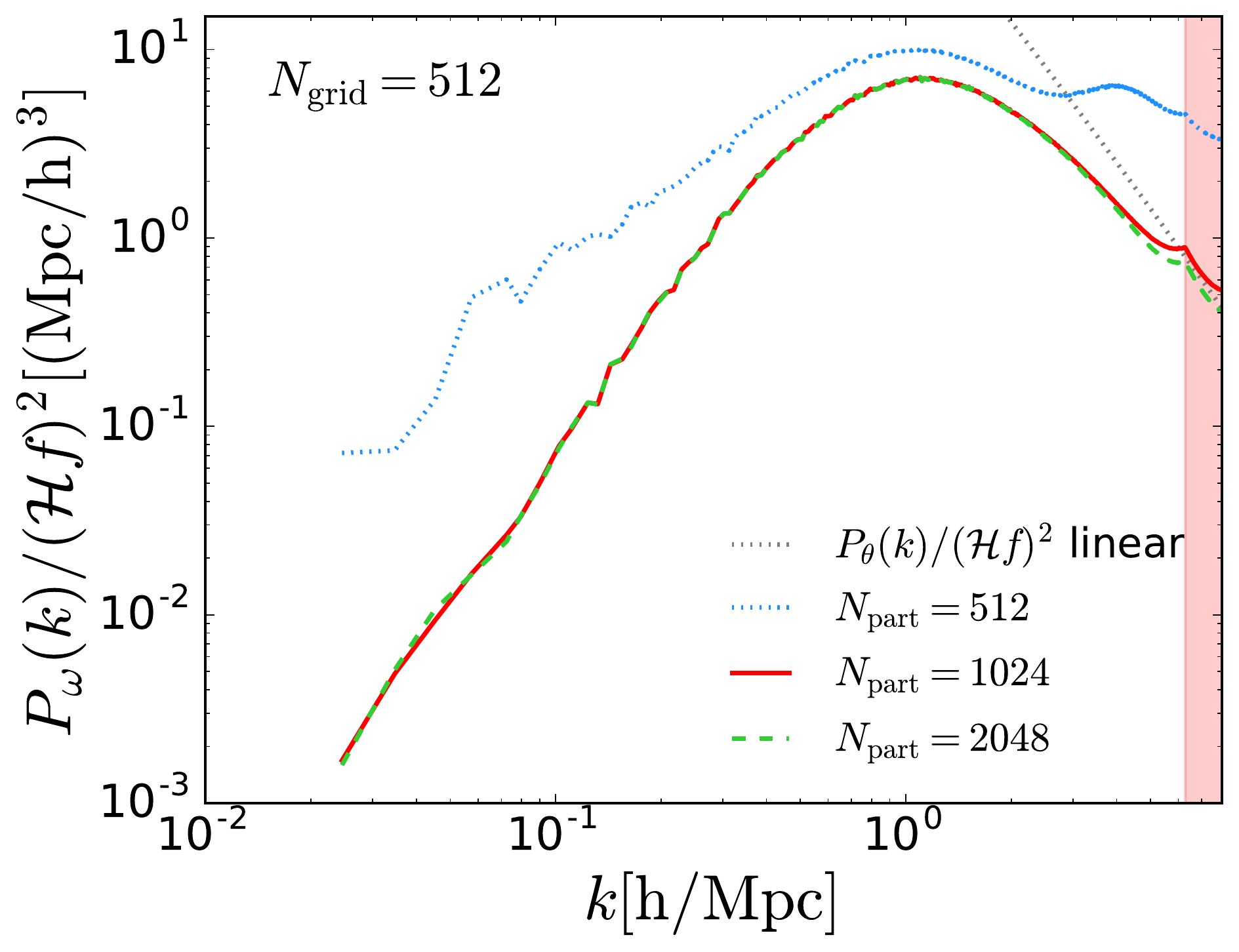}
\caption{The vorticity power spectrum for $R=1,2,4$ at $z=0$, with $N_\text{grid}=512$ and $L=256$ Mpc/$h$}
\label{fig:part_per_grid}
\end{figure}

The results are displayed in figure \ref{fig:part_per_grid}.
As we can immediately see, the vorticity power spectrum is significantly higher for the case $R=1$ than for the others.
On the other hand, the values $R=2$ and $R=4$ give nearly identical results, except at the smallest scales close to the Nyquist frequency.
As the $R=4$ simulations would be much more computationally demanding, in almost all of our simulation runs we have used the value $R=2$, corresponding to 8 particles per grid cell on average.

\subsubsection{Dependence on spatial resolution}
\label{s:converg2}
In order to make sure the results are reliable, we have tested the dependence of the obtained power spectra on the spatial resolution of the simulation.
In figure \ref{fig:ps_res_test} we show the velocity divergence (left panels) and vorticity (right panels) power spectra for the
three simulations in table \ref{Table-sims}, which have number of particles per side $N_\text{part}=1024, 2048, 4096$ with a fixed particle-to-grid ratio $N_\text{part}/N_\text{grid}=2$. Top and bottom panels refer to redshifts $z=0$ and $z=1$, respectively.

\begin{figure}[h!]
    \centering
    \begin{subfigure}[b]{0.495\textwidth}
        \includegraphics[width=\textwidth]{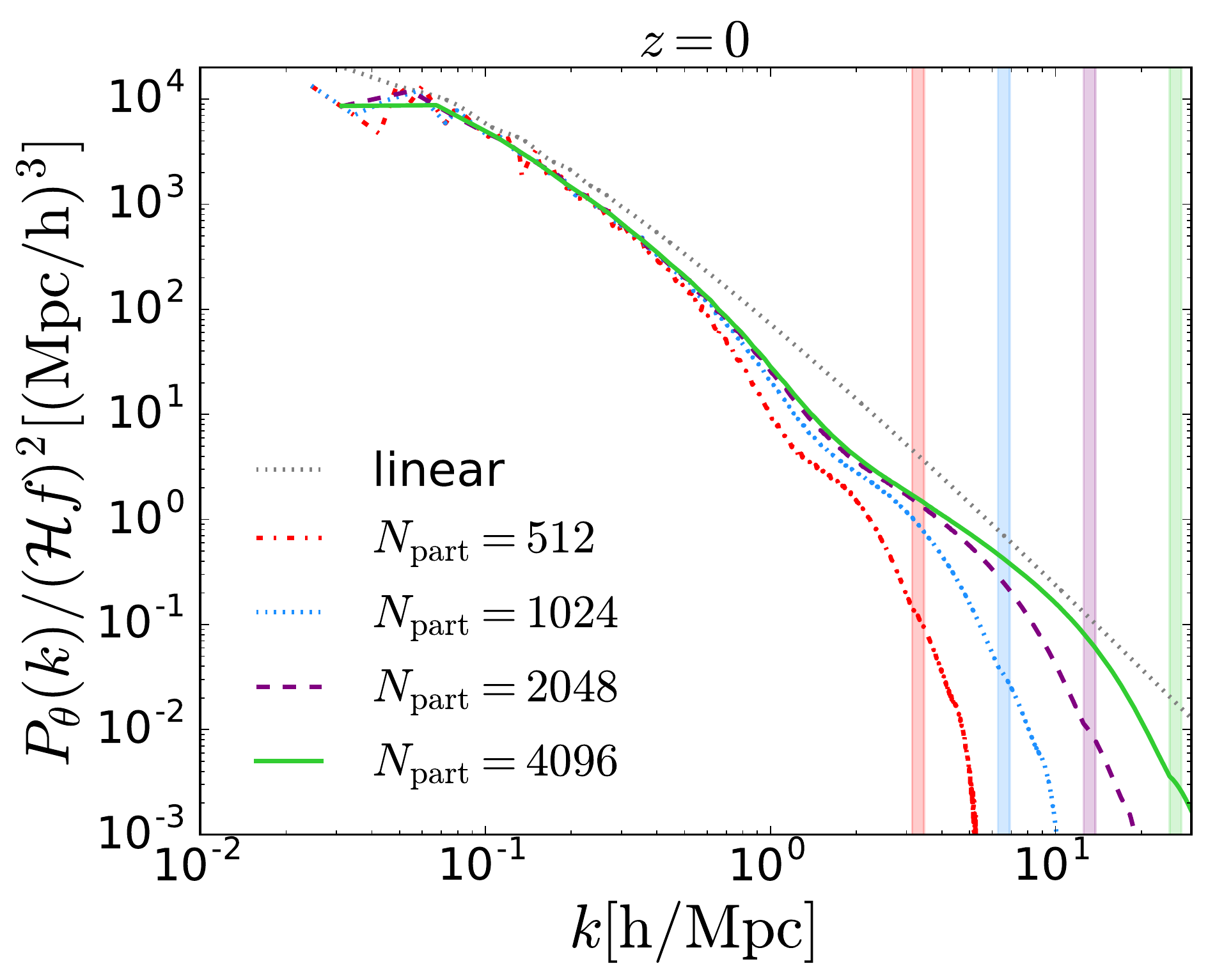}
        \label{smooth:th0}
    \end{subfigure}
    \begin{subfigure}[b]{0.495\textwidth}
        \includegraphics[width=\textwidth]{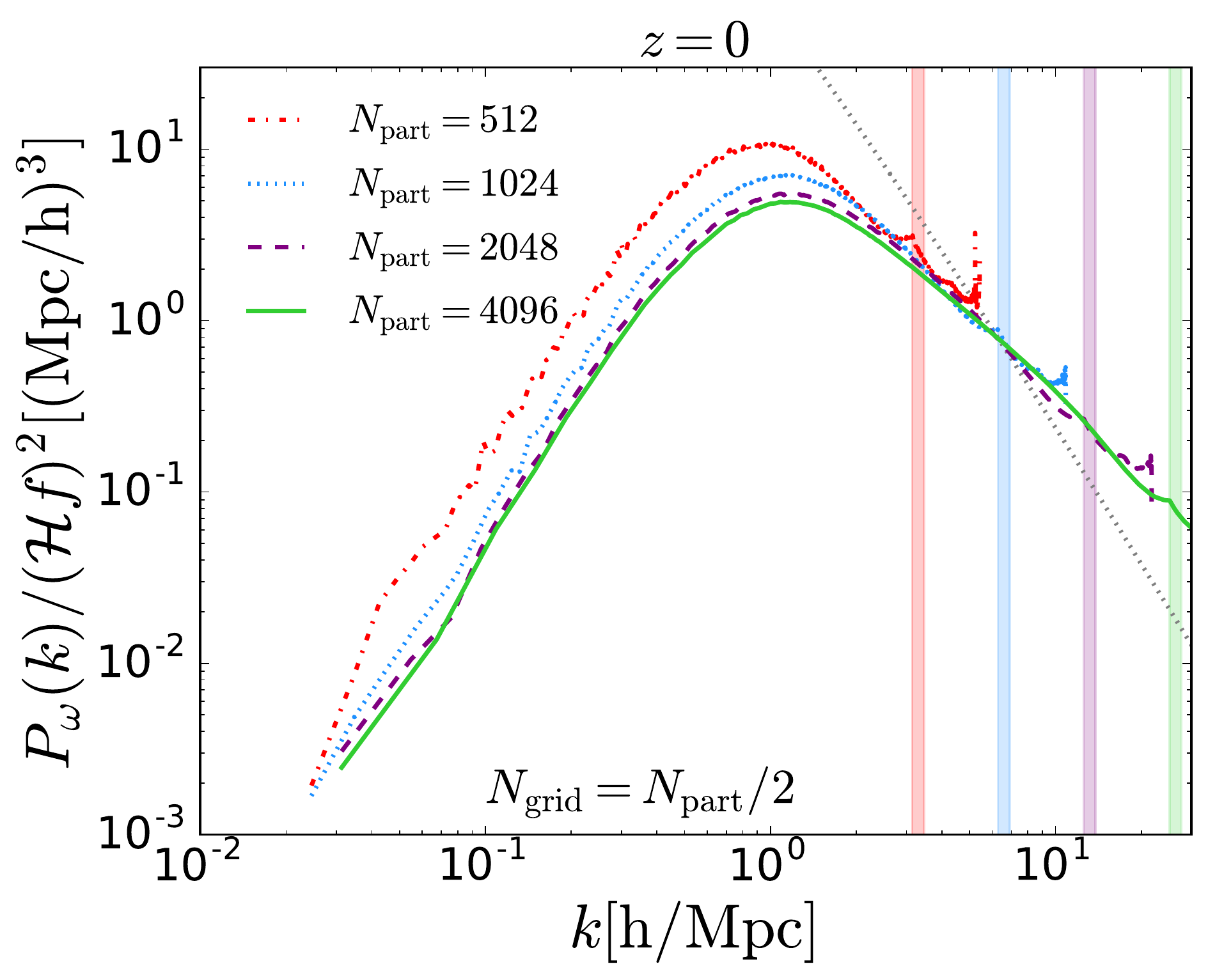}
        \label{smooth:w0}
    \end{subfigure}
    \\
    \begin{subfigure}[b]{0.495\textwidth}
        \includegraphics[width=\textwidth]{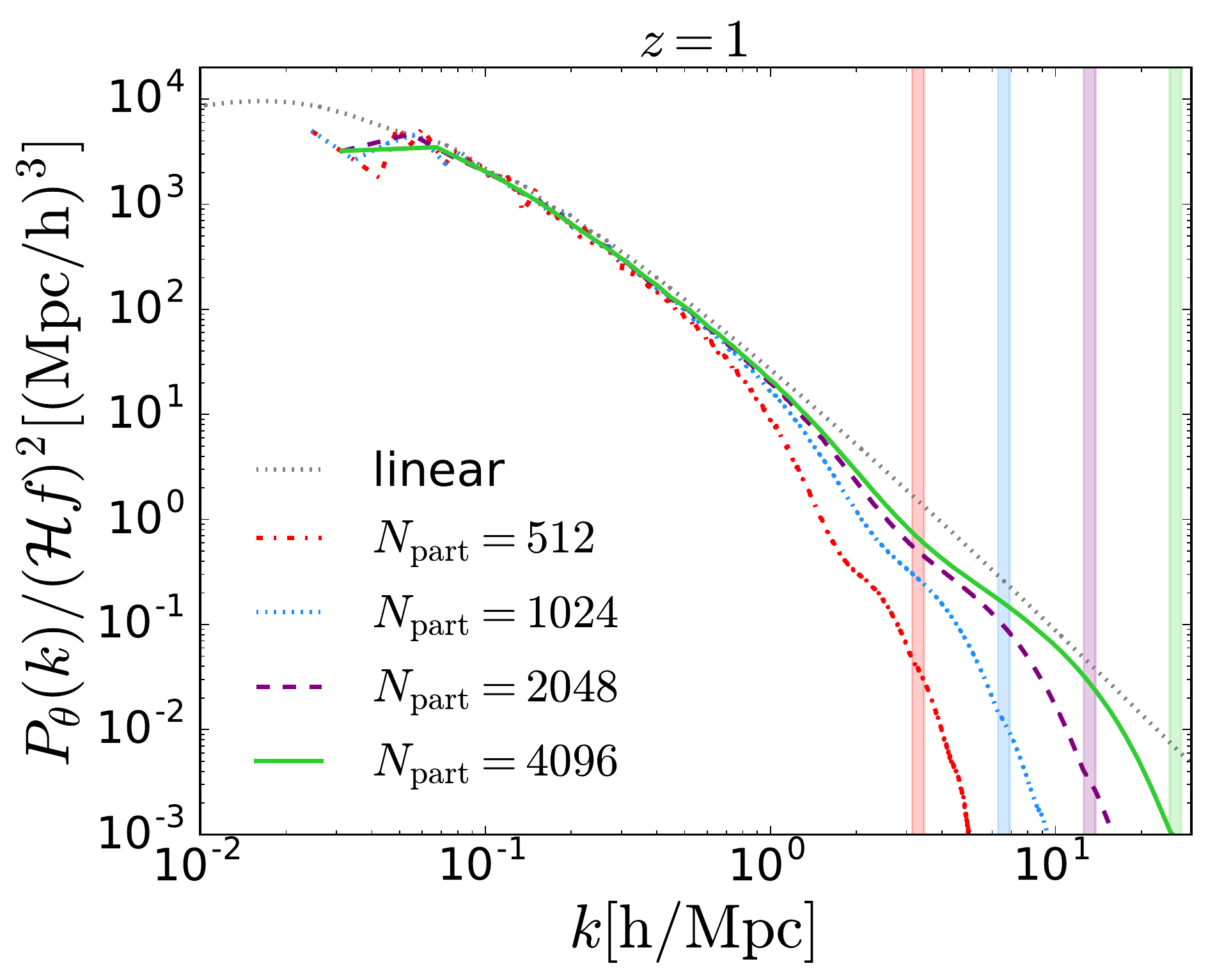} 
        \caption{Power spectrum - $\theta$}
        \label{smooth:th1}
    \end{subfigure}
    \begin{subfigure}[b]{0.495\textwidth}
        \includegraphics[width=\textwidth]{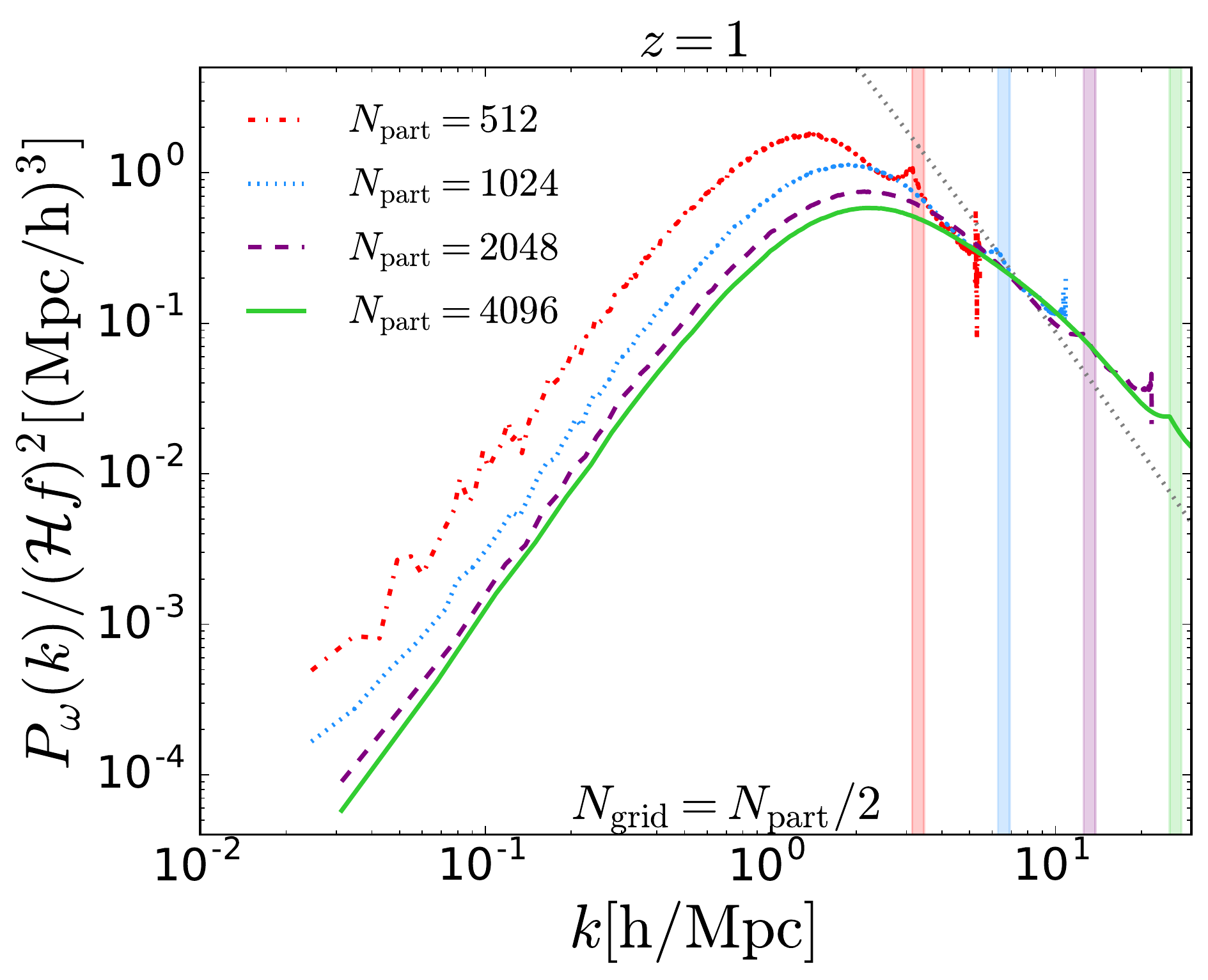}
        \caption{Power spectrum - $\omega$}
        \label{smooth:w1}
    \end{subfigure}
    \caption{Velocity divergence and vorticity power spectra at $z=0$ (\textit{top}) and $z=1$ (\textit{bottom}) respectively. The highlighted bands represent the Nyquist frequency for each resolution.}
    \label{fig:ps_res_test}
\end{figure}
\iffalse
\begin{figure}[H]
\centering
\includegraphics[scale=0.3]{pkwz0.pdf}
\caption{Velocity divergence and vorticity power spectra at $z=1$ (\textit{top}) and $z=0$ (\textit{bottom}) respectively. The highlighted bands represent the Nyquist frequency for each resolution.}
\label{fig:ps_res_test}
\end{figure}
\fi

We notice that, at large scales, the velocity divergence spectrum is not very sensitive to the spatial resolution, while the dependence increases as we go to nonlinear scales. Of course we can expect the simulations to agree only on scales which are sufficiently larger than their Nyquist frequencies.

On the other hand, the power spectrum of the vorticity, being an entirely nonlinear phenomenon, exhibits a resolution dependence at all scales, notably the amplitude is too high for low resolution simulations, and it eventually converges for a sufficient spatial resolution, while the shape of the spectrum is roughly resolution independent.

Comparing the $z=0$ and $z=1$ plots, we see that the convergence is seemingly worse for higher redshift.
This can be attributed to the fact that shell (orbit) crossing, which is the generation mechanism for vorticity in our simulation, occurs much less at higher redshifts.
Therefore, the spurious vorticity induced by numerical velocity dispersion resulting from the CIC projection of the particle velocities on the grid is more significant when there is less physical vorticity in our simulation, and we would need a higher spatial resolution to obtain reliable results at higher redshifts.
Consequently, we expect the power spectrum of the vorticity to have a progressively worse convergence as we go to higher redshifts. 

\begin{figure}[t]
\begin{center}
\includegraphics[scale=0.5]{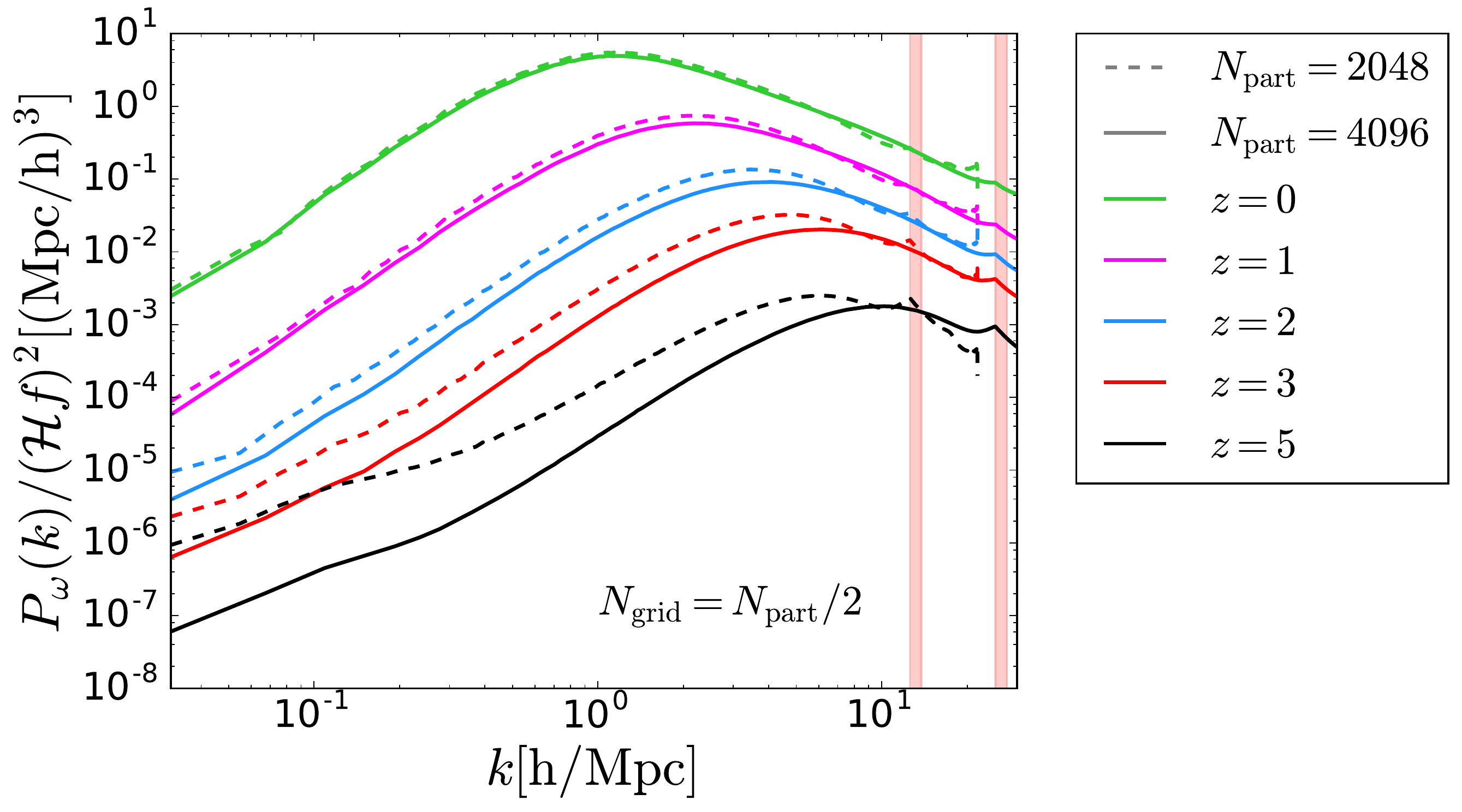}
\caption{The vorticity power spectra for redshifts $z=0,1,2,3,5$.}
\label{fig:pw_redshift}
\end{center}
\end{figure}

This is shown in figure \ref{fig:pw_redshift}, where we compared the two simulations \textit{high resolution 1} and \textit{high resolution 2} in table \ref{Table-sims} at redshift $z = 0, 1, 2, 3, 5$.
While for $z=0$ and $z=1$ the vorticity power spectra are in good agreement, for $z > 1$ the difference between the results of the two simulations is significant and this trend worsens with increasing redshift. 
For this reason, we focus our attention primarily on the redshift range from $z=0$ to $z=1$.

It is interesting to notice in figure \ref{fig:pw_redshift} that the position of the peak of the vorticity power spectrum is redshift dependent, and it is moving from small scales at high redshift to larger scales at $z = 0$. 
Between $z = 1$ and $z = 0$ the peak position is shifted from $k_\text{p} \approx 2$ $h$/Mpc to $k_\text{p} \approx 1$ $h$/Mpc, while its amplitude grows from $P_\om/(\HH f)^2(k_\text{p}, z = 1) \approx 0.6 \, \text{Mpc}^3/h^3$ to 
$P_\om/(\HH f)^2(k_\text{p}, z = 0) \approx 5 \, \text{Mpc}^3/h^3$.
This shift of the peak is roughly in the same position as the dip in the divergence power spectrum, see figure~\ref{fig:ps_res_test}, and it indicates the scale where non-linearities become strong which moves to successively larger scales.
As mentioned in the previous subsection, we interpret this as the scale at which further infall of particles is slowed down due to angular momentum conservation.
Similar results have also been obtained in \citep{zhu1, zhu2}.

Following \citep{Pueblas:2008uv}, we have also considered in our analysis possible aliasing effects due to the finite size of the grid and the use of discrete Fourier transforms, however they were found to be negligible for our resolution and redshift range.

\subsubsection{Redshift dependence}
\label{s:redshift}
\begin{figure}[t]
\begin{center}
\includegraphics[width=0.6\textwidth]{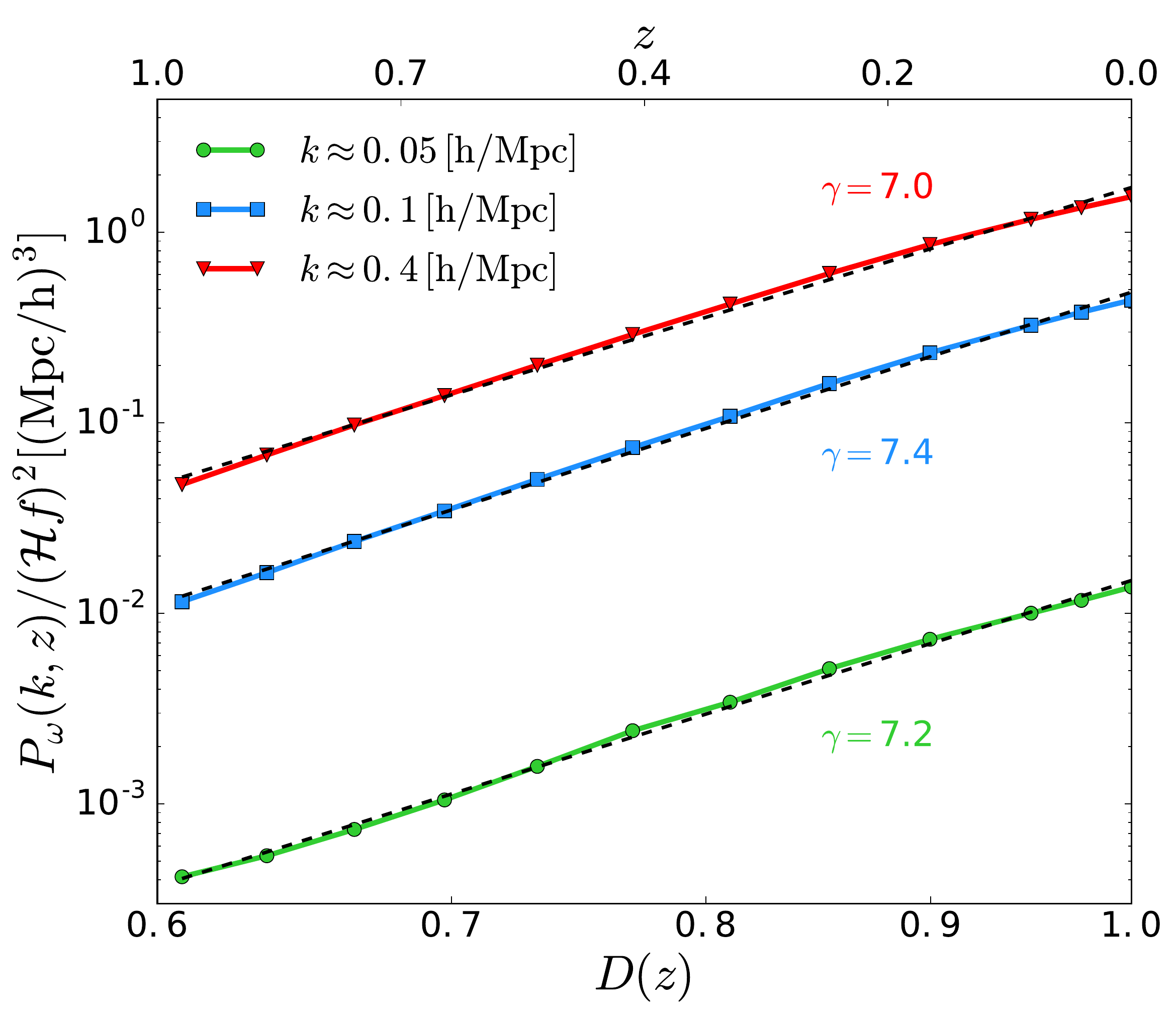}
\caption{
Vorticity power spectrum at fixed mode $k$ as a function of the growth function (bottom axis) and redshift (top axis). The black dashed lines are the fit to the numerical data.
}
\label{f:PDz}
\end{center}
\end{figure}

Focusing primarily on the range $z\in[0,1]$, we have studied the large scale redshift dependence of the vorticity power spectrum.
We have chosen three different modes in the large scale regime, $k=0.05, 0.2, 0.4$ $h$/Mpc.
Figure \ref{f:PDz} shows the redshift and growth-factor dependence of the vorticity power spectrum at fixed scale.
Following \citep{Pueblas:2008uv}, we have assumed that the vorticity power spectrum has a redshift dependence of the form 
\be P_\omega(k,z)\propto (\mathcal{H}f)^2 D^\gamma(z). \label{PDz}
\ee 
We have performed a least squares fit to the data; the fits to the numerical data are displayed in black dashed colors in figure \ref{f:PDz}.
We found values of $\gamma$ in the range $[7, 7.4]$, which is mostly consistent with the result obtained in \cite{Pueblas:2008uv, Thomas:2015kua}.
However, we note that a) we are not able to test the fit (\ref{PDz}) for higher redshift due to the convergence issued outlined in the previous section, b) we cannot obtain a good fit on all the scales of the simulations with a global parameter $\ga$. Our findings on the redshift evolution of the vorticity spectrum confirm the results of \cite{Thomas:2015kua}, namely that the growth factor for the vorticity spectrum cannot be modelled as a simple power law in a wide redshift range.

We also notice that the time evolution of the vorticity spectrum extracted from N-body simulation results is very different from the result obtained from a perturbative approach.
In fact, the power spectrum computed in \cite{Cusin:2016zvu} grows linearly with the growth function, 
\be
P^\text{pert}_\omega(k,z)\propto D(z).
\ee
As already noted in the previous section, this very significant difference in the redshift evolution is probably due to the different physical mechanism that is generating vorticity in the two cases. Being a perturbative approach, the method discussed in~\cite{Cusin:2016zvu} cannot really handle shell crossing which is the dominant effect leading to vorticity production in N-body simulations, as we shall see below.

\begin{figure}[h!]
\centering
\begin{minipage}{.45\textwidth}
\centering
\includegraphics[width=\linewidth]{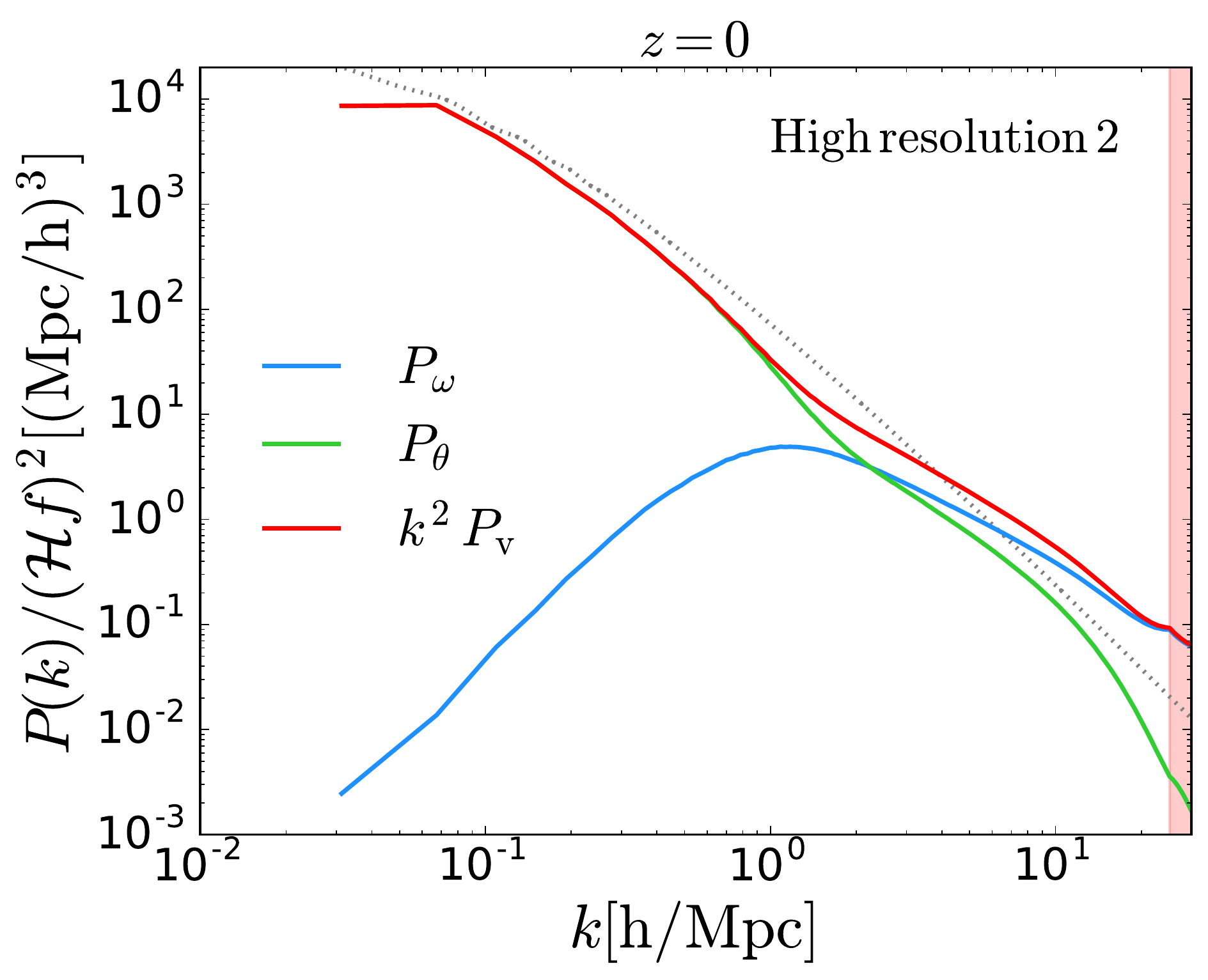}
\end{minipage}
\begin{minipage}{.45\textwidth}
\centering
\includegraphics[width=\linewidth]{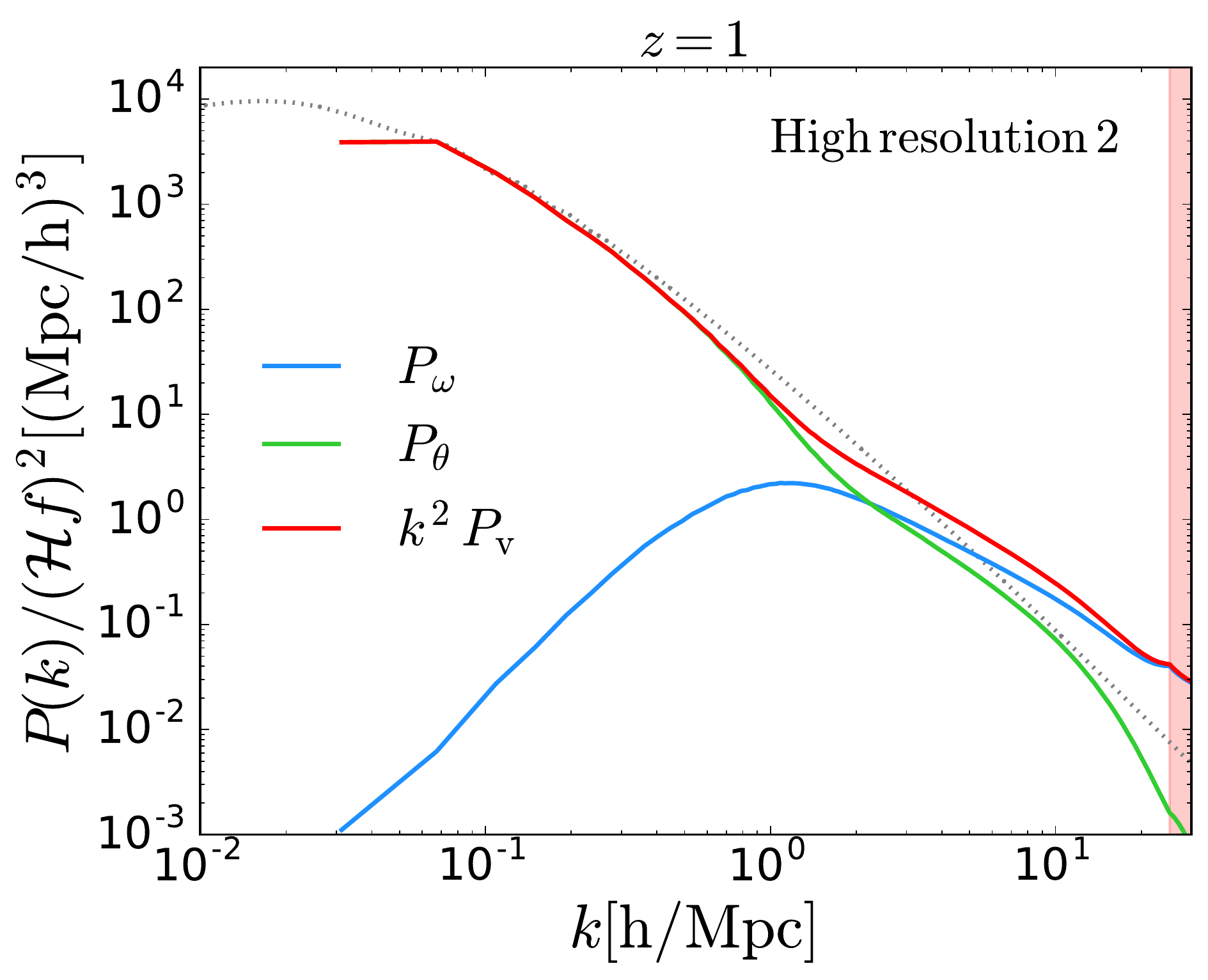}
\end{minipage}
\caption{Vorticity, divergence and velocity power spectra at $z=0$ (left panel) and $z=1$ (right panel). Both plots refers to the
simulation \emph{high resolution 2} in table \ref{Table-sims}. The dotted line represents the linear divergence spectrum.}
\label{fig:k2Pv}
\end{figure}

In figure \ref{fig:k2Pv} we plot the divergence and vorticity spectra together with the total velocity power spectrum $k^2 P_\text{v} = P_\omega + P_\theta$ at $z = 0$ (left panel) and $z=1$ (right panel). On large scales, as expected, the divergence contributes most to the total velocity spectrum. However, on small scales the divergence spectrum drops faster than the vorticity, and, beyond $k \approx 2$ $h$/Mpc, where the velocity power spectrum exceeds the linear perturbation theory result, vorticity becomes the dominant component in shaping the total velocity power spectrum. 

\subsection{Cross-spectra}
\label{s:cross}

In our simulations we have also computed the cross-spectra of the velocity divergence with the density contrast and the velocity dispersion $\sigma^2$, and the cross-correlation of the rotational part of the velocity field with the metric vector perturbations $\mathbf{B}$ \footnote{Note that the cross-spectrum $P_{\omega B}$ vanishes by parity conservation; under spatial inversions, $\omega$ is a pseudovector, while $B$ is a true vector, and the relationship between the correlators in Fourier space is $\langle \omega_i B^*_j\rangle = i\epsilon_{imn} k_m \langle v_{R n}B^*_j\rangle$, which would be odd under parity, and hence must be zero.}.
The results of this section refer to the simulation \emph{high resolution 1} in table \ref{Table-sims}.

\begin{figure}[h!]
\centering
\begin{minipage}{.45\textwidth}
\centering
\includegraphics[width=\linewidth]{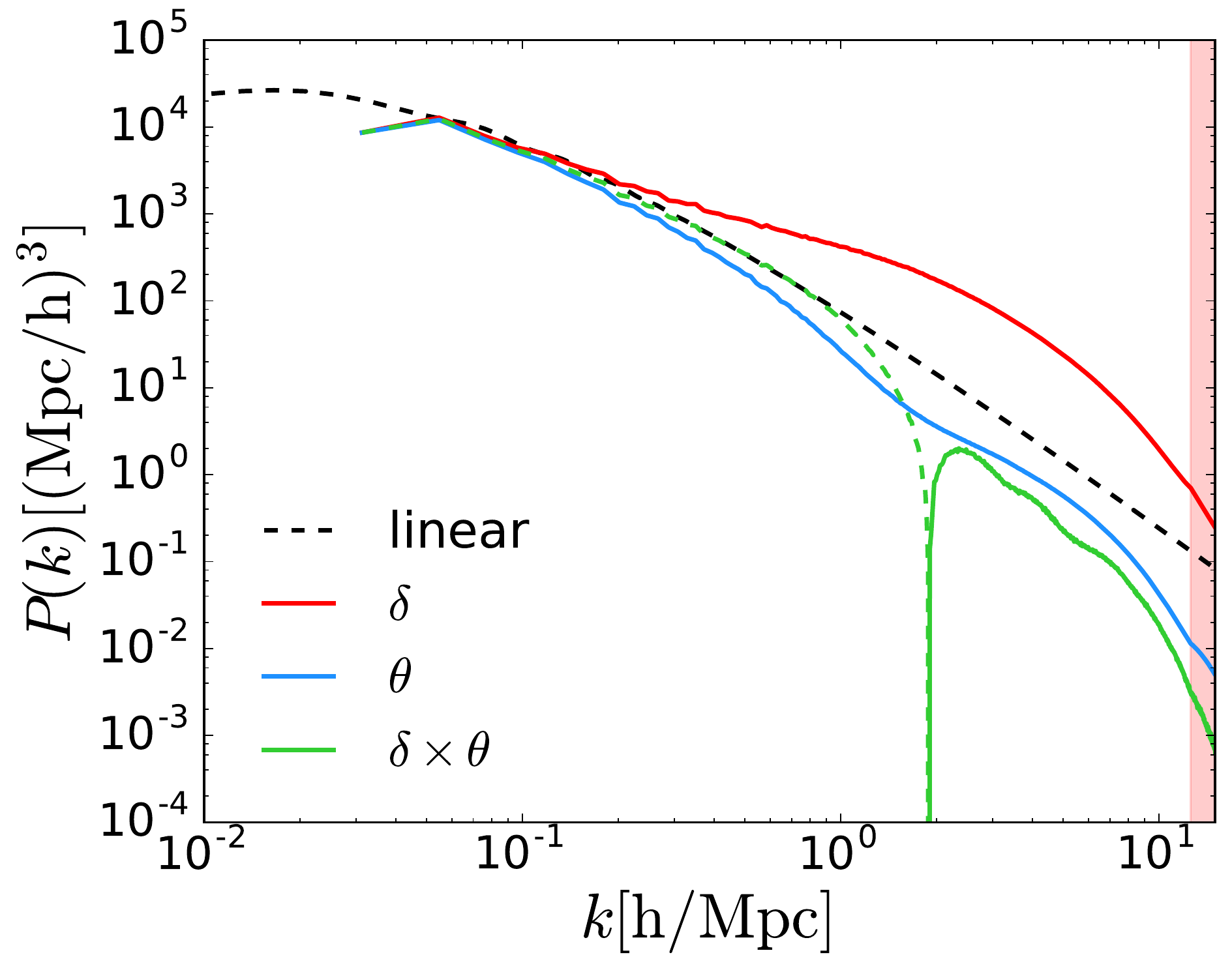}
\end{minipage}
\begin{minipage}{.45\textwidth}
\centering
\includegraphics[width=\linewidth]{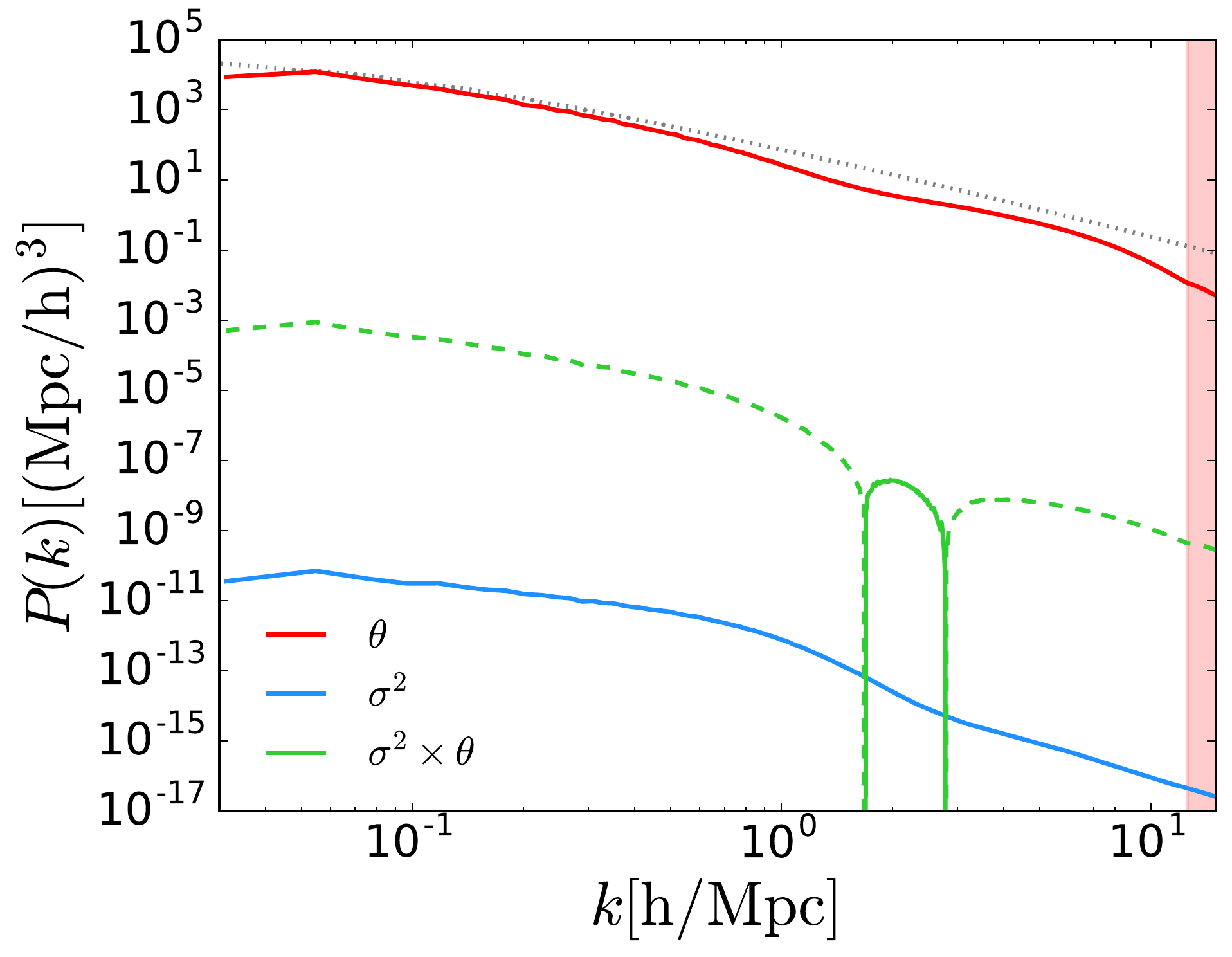}
\end{minipage}
\caption{Left panel: The spectra of the density contrast (red), the velocity divergence (blue), and their cross-spectrum (green) at $z=0$. Right panel: The spectra of the velocity divergence (red), the velocity dispersion (blue), and their cross-spectrum (green) at $z=0$.}
\label{fig:spectra_theta}
\end{figure}

In figure \ref{fig:spectra_theta} (left panel) we plot the power spectra of the density contrast, velocity divergence, and their cross spectra, respectively, at $z=0$. Both the power spectrum of the divergence and the cross-spectrum are normalized in such a way that all the spectra are dimensionally consistent. From the linearized continuity equation we expect on large scales $\theta =- {\mathcal{H}f} \delta$.
Therefore, we normalize the divergence spectrum by a factor $(\HH f)^2$ and the cross-spectrum by a factor $\HH f$.
Consistent with these definitions, on large scale all the spectra are approximately equal. On the other hand, due to non-linear evolution at late times, the divergence spectrum drops on smaller scales, while the density spectrum increases. Interestingly, the cross correlation spectrum remains close to the linear perturbation theory value until $k\simeq 1h/$Mpc where it increases, until it changes sign at $k \approx 2$ $h$/Mpc and then settles at a fraction of the divergence spectrum.
This is an indication of the scale at which shell-crossing is relevant. In fact, after shell-crossing, which leads to the formation of structures, the initially anti-correlated density and velocity divergence are positively correlated on small scales. We understand that after shell crossing a previous infall onto a filament changes sign becomes actually an outflow, leading to positive density $-$ divergence correlations. See also \cite{Hahn:2014lca} for a discussion of this.

\begin{figure}[h!]
\centering
\begin{minipage}{.45\textwidth}
\centering
\includegraphics[width=\linewidth]{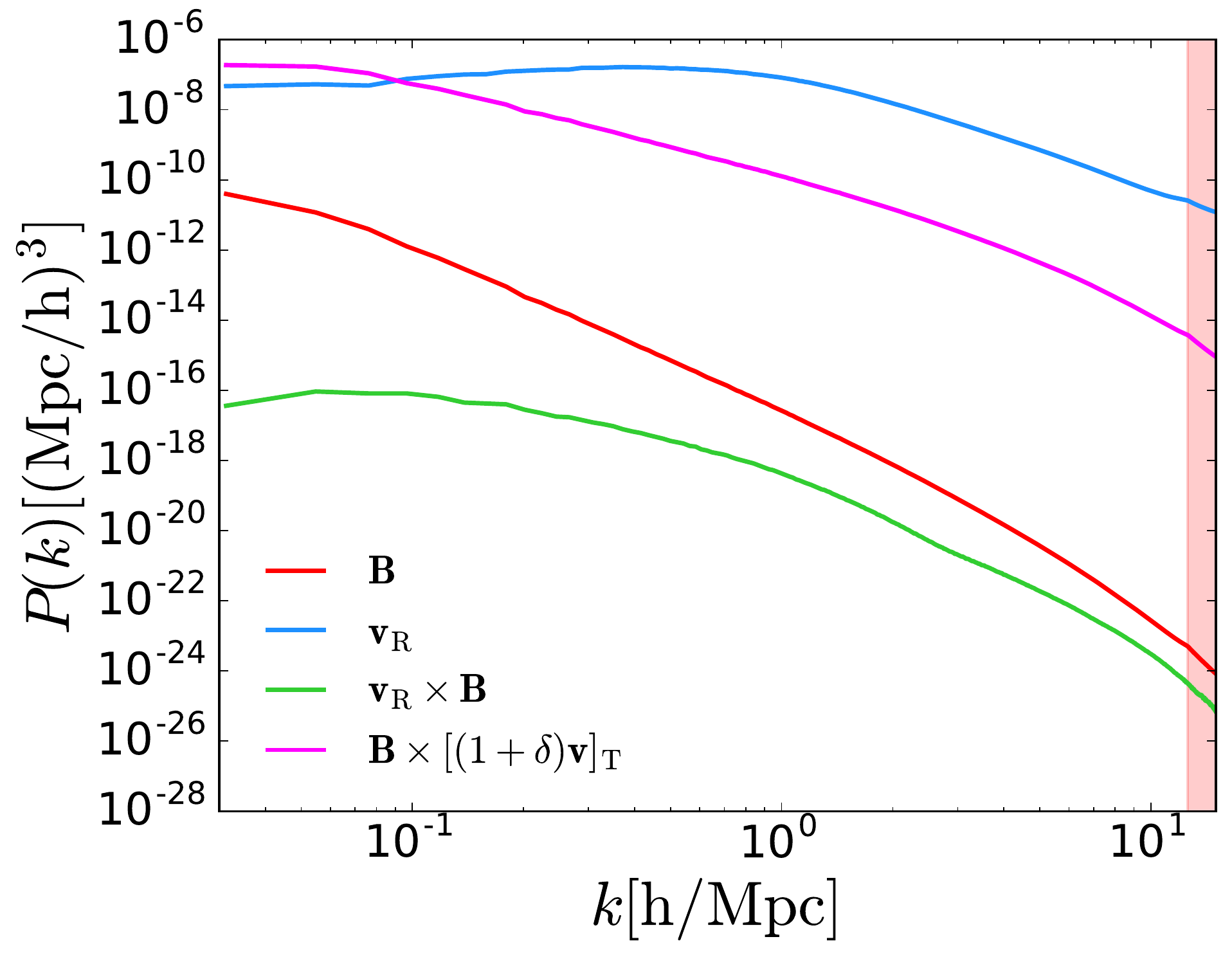}
\end{minipage}
\begin{minipage}{.45\textwidth}
\centering
\includegraphics[width=\linewidth]{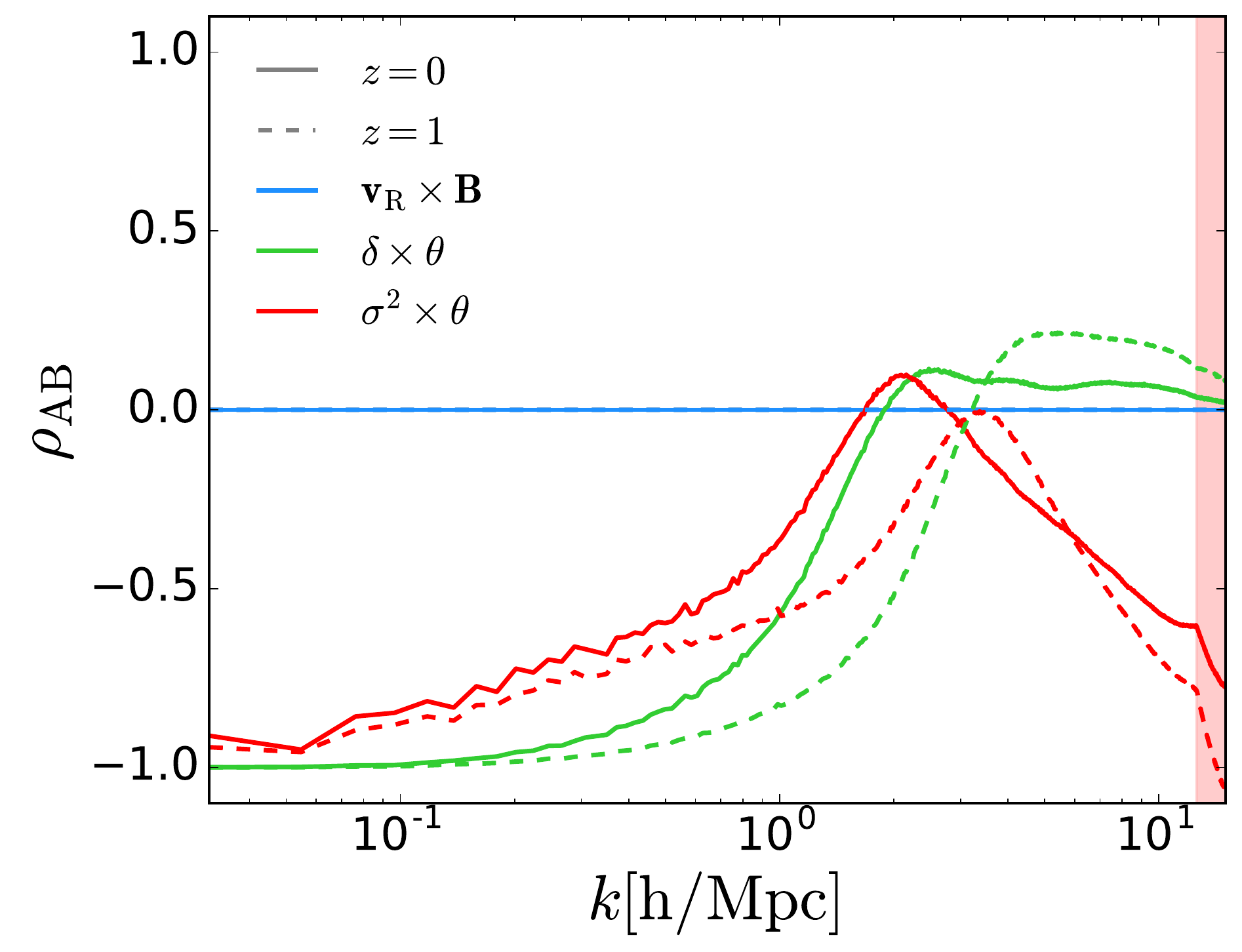}
\end{minipage}
\caption{Left panel: The spectra of the metric vector perturbations (red), the rotational velocity (blue), and their cross-spectrum (green) at $z=0$. The magenta line represents the cross-correlation of the metric perturbation with its source term (see equation \eqref{e:BvR}). Right panel: The correlation coefficients between the rotational velocity $v_R$ and the metric vector perturbation $B$ (blue), the density contrast $\delta$ and velocity divergence $\theta$ (green), the velocity dispersion $\sigma$ and velocity divergence $\theta$ (red), at $z=0$ (solid) and $z=1$ (dashed) respectively.}
\label{fig:spectra_B}
\end{figure}

In figure \ref{fig:spectra_theta} (right panel) we display the power spectra of the velocity divergence, the velocity dispersion, and their cross spectra, respectively, at $z=0$. 
On large scales, the velocity dispersion spectrum is proportional to the $\theta$ spectrum and the two are strongly anti-correlated.
This is because the fluctuation of particle velocities around their mean bulk flow is a second order effect that is most efficiently generated in collapsing regions which have negative $\theta$.
However, at small scales, the dependence becomes highly non-linear and the cross correlation between $\sigma^2$ and $\theta$ even briefly changes sign around the scale where also $\de\time\theta$ exhibits a sign change.
As one can see in figure~\ref{fig:spectra_B}, at $z=1$ there is no sign change in the correlation of $\si^ 2$ and $\theta$, and they remain anti-correlated on all scales.

In figure \ref{fig:spectra_B} (left panel) we plot the power spectra for $\mathbf{B}$ and $\mathbf{v}_R$ and their cross spectrum.
The rotational part of the velocity is expected to partially source the vector perturbation in the metric. 
In fact, by taking the transverse projection of the $0i$ Einstein constraint equation linearized in $\mathbf{B}$, we find
\begin{equation}
k^2 \mathbf{B} = 6 \HH^2 \Omega_\text{m} [(1+\delta)\mathbf{v}]_\text{T} =6 \HH^2 \Omega_m [ \mathbf{v}_R +(\delta\,\mathbf{v})_\text{T}] ,
\label{e:BvR}
\end{equation}
where the subscript T denotes the transverse projection of the source term $(1+\delta)\mathbf{v}$.
In figure \ref{fig:spectra_B} we compare the cross-spectrum of the metric perturbations $\mathbf{B}$ with the rotational
velocity (green line), and with the full source term in equation \eqref{e:BvR}, which is computed from equation
\eqref{e:BvR}, i.e. $P_{\text{B}\, [(1+\delta) \text{v}]_\text{T}} = P_{\text{B}}\, k^2 /(6 \HH^2 \Omega_\text{m})$.
The cross-spectrum of $\mathbf{B}$ with $\mathbf{v}_R$ is smaller by many orders of magnitude than
the cross-spectrum of $\mathbf{B}$ with the full source term. Therefore, we expect that the main source of vector 
perturbations does not come from the rotational velocity, but from the product of the velocity with the density contrast, i.e. from
$(\delta\,\mathbf{v})_\text{T}$. This also agrees with the finding of~\cite{Thomas:2015kua}, see their figure~7. There they find that $\nabla\times \bB$ is mainly generated by $\nabla\de\times \bv$, and the contribution from vorticity is many orders of magnitude smaller. The B-power spectra from both simulations also agree qualitatively, see figure~4 of~\cite{Thomas:2015kua}. Note that in this reference, the dimensionless power spectra are shown which differ from ours by a factor $k^3$.

We now define the correlation coefficient for two generic quantities 
A and B as
\be
\rho_\text{AB}\equiv \frac{P_\text{AB}}{\sqrt{P_\text{A} P_\text{B}}} \,.
\ee

The correlation coefficients for the rotational velocity $\mathbf{v}_R$ and the metric vector perturbation $\mathbf{B}$, the density contrast $\delta$ and velocity divergence $\theta$, and the velocity dispersion $\sigma^2$ and velocity divergence $\theta$ are displayed in figure \ref{fig:spectra_B} (right panel) at $z = 0$ and $z = 1$.

As discussed before, both the density contrast and the velocity dispersion are anti-correlated with $\theta$ on large scales.
We also notice that the scale at which the correlation coefficients for $\delta \times \theta$ becomes positive depends on redshift.
This is due to the fact that the regions that experience shell-crossing are larger at low redshift and, therefore, the velocity divergence
becomes positive at larger scales. The sign change of the $\theta\times\delta$ and $\theta\times\si^2$ cross spectra are roughly at the same scale which is somewhat smaller than the scale of the peak of the vorticity spectrum. Below this scale density and divergence are weakly correlated and positive which indicates a slight net outflow instead of inflow around structures. On large scales, the velocity dispersion is largest where the density is highest, hence clearly also $\si^2$ and $\theta$ are anti-correlated. This anti-correlation nearly vanishes and for $z=0$ even becomes slightly positive on the scales where further inflow is prohibited by angular momentum conservation.

The correlation coefficient between $\mathbf{B}$ and the rotational velocity is roughly $10^{-7}$ on all scales and therefore not visible on this linear plot. As discussed before,
this means that the main source for the metric perturbation does not come from the rotational velocity alone but
from the transverse part of $(\delta\,\mathbf{v})$.

\section{Conclusions}
\label{s:con}

In this work we numerically investigate the generation of vorticity from a set of N-body simulations.
We have implemented the computation of the velocity field and its dispersion in the relativistic N-body code \emph{gevolution}.
The velocity field is estimated from the momentum and density fields, which are computed in \emph{gevolution} using a particle-to-mesh projection.
In order to validate our method, we have tested several prescriptions to deal with the empty regions of the simulations: an identically zero velocity field in voids; a Gaussian smoothing for the density and momentum field, from which the velocity is computed; a velocity field which does not evolve in voids and a velocity field which evolves in the empty regions according to the linear growth rate.
All these methods give very similar results for the vorticity power spectrum.
However, the main results presented in this paper are obtained with the latter method, which is also physically well motivated. 

The key findings of this work can be summarized as follows:
\begin{enumerate}[label=(\alph*)] 
\item On large sub-horizon scales, the vorticity power spectrum is well modelled by a power-law with a spectral index $n_\omega \approx 2.5$.
This result quantitatively agrees with the analysis presented in~\cite{Hahn:2014lca}.
Causality actually requires $n_\om=2$ on the largest scales, but we could not see this in our simulations. % We suppose that this is a finite size effect.
In fact, we expect some change of slope around the horizon scale which, however, is not represented in our simulations because of the choice of box size.
\item At redshift $z=0$ the vorticity power spectrum peaks at $k_p(z=0)\simeq 1$ $\textit{h}$/Mpc.
At roughly this scale also the divergence power spectrum has a dip and the density $-$ divergence correlation changes sign, a clear indication of shell crossing. The scale $k^{-1}_p(z)$ decays with increasing redshift to $k_p(z=1)\simeq 2$ $\textit{h}$/Mpc.
\item The vorticity power spectrum in our simulations depends both on the mass resolution (number of particles per cell) and grid resolution (number of grid points).
We need at least 8 particles per grid cell and a grid resolution $\Delta x = 125$ kpc/$h$ in order for the vorticity power spectrum to converge at $z \in [0,1]$. 
\item The grid resolution that is needed in order to have convergence depends on redshift. At higher redshift, shell-crossing is a rarer phenomenon than at $z=0$, and it happens on smaller scales.
Therefore, a finer grid resolution is required in order to resolve shell-crossing and disentangle it from numerical velocity dispersion that is always present in our simulations due to finite resolution.
\item In the redshift range $z \in [0, 1]$, where our results are reliable, the vorticity power spectrum grows as $(\HH f)^2 D^{\gamma}(z)$, with $\gamma \approx 7$.
This result is consistent with \cite{Pueblas:2008uv, Thomas:2015kua}.
\item The amplitude of the vorticity spectrum 
at the peak position increases from\\ $P_\omega/(\HH f)^2(k_p) \approx 0.6\, (\text{Mpc}/\textit{h})^3$ at $z = 1$ to 
$P_\omega/(\HH f)^2(k_p)  \approx 5\, (\text{Mpc}/\textit{h})^3$ at $z = 0$.
Roughly at this peak position the divergence power spectrum drops significantly below the linear perturbation theory result and the vorticity becomes the dominant contributor to the total velocity spectrum.
\item The rotational part of the velocity is very weakly correlated with the vector perturbations of the metric.
Therefore, we expect the metric perturbations $\mathbf{B}$ to be predominantly sourced by the combination of $\delta$ and the velocity field, $(\delta\bv)_T$ and not by the velocity field alone. 
\end{enumerate}

Observationally, the presence of vorticity will have interesting consequences e.g. for redshift space distortions. Our results indicate that the best place to look for them is around the Mpc scale which is also the scale where we expect predictions from linear theory to become unreliable. It therefore remains a challenge to tackle this problem, and numerical simulations will probably continue to play an important role in this context.

\subsection*{Acknowledgements}
It is a pleasure to thank David Daverio, Andreas Finke, Vittorio Tansella, and Cornelius Rampf for useful discussions and comments on the manuscript.
We also thank Daniel B. Thomas for help with the publicly available DTFE code.
We are grateful to the anonymous referee for the constructive feedback.
FL thanks the University of Geneva for hospitality.
This work was supported by a grant from the Swiss National Supercomputing Centre (CSCS) under project ID s710.
The numerical simulations were carried out on \textit{Piz Daint} at the CSCS and on
the \textit{Baobab} cluster of the University of Geneva. FL is supported by the INFN grant PD 51 INDARK. RD and GJ are supported by
the Swiss National Science Foundation. 

\subsection*{Disclaimer}
This is an author-created, un-copyedited version of an article published in the Journal of Cosmology and Astroparticle Physics (JCAP). IOP Publishing Ltd is not responsible for any errors or omissions in this version of the manuscript or any version derived from it. The Version of Record is available online at \url{https://doi.org/10.1088/1475-7516/2018/09/006}

\vspace{1.5cm}
\appendix
\noindent{\LARGE \bf Appendix}

\section{Finite-box effect}
\label{a:finbox}
In this section we test the effect of the finite box employed for our simulations on our power spectra.

We run two simulations with the same spatial resolution and different box size. Their parameters are summarized in
table \ref{Table-sims3}. Both SIM1 and BIGSIM1 adopt the \textit{rescaled} method to extract the 
velocity field. The physical box size $L$ of BIGSIM1 is twice larger than the one of SIM1.

\begin{table}[h!]
\centering
\begin{tabular}{|c|c|c|c|c|}
\hline
\multicolumn{1}{|c|}{}                                           & 
 \multicolumn{1}{c|}{$N_\text{grid}$}                   &
  \multicolumn{1}{c|}{$N_\mathrm{part}$}                      &
  \multicolumn{1}{c|}{\textit{L} [Mpc/\textit{h}]}                        &
  \multicolumn{1}{c|}{Grid-Cell size [Mpc/\textit{h}]}                         \\
\hline
SIM1            &    512    &  1024       &    256      &   0.5     \\
BIGSIM1      &   1024    &  2048     &     512     &    0.5      \\
\hline
\end{tabular}
\caption{
Summary of the simulations used for testing the effect of a finite box.
}
\label{Table-sims3}
\end{table}

\begin{figure}[h!]
\centering
\begin{subfigure}[b]{0.495\textwidth}
\includegraphics[width=\textwidth]{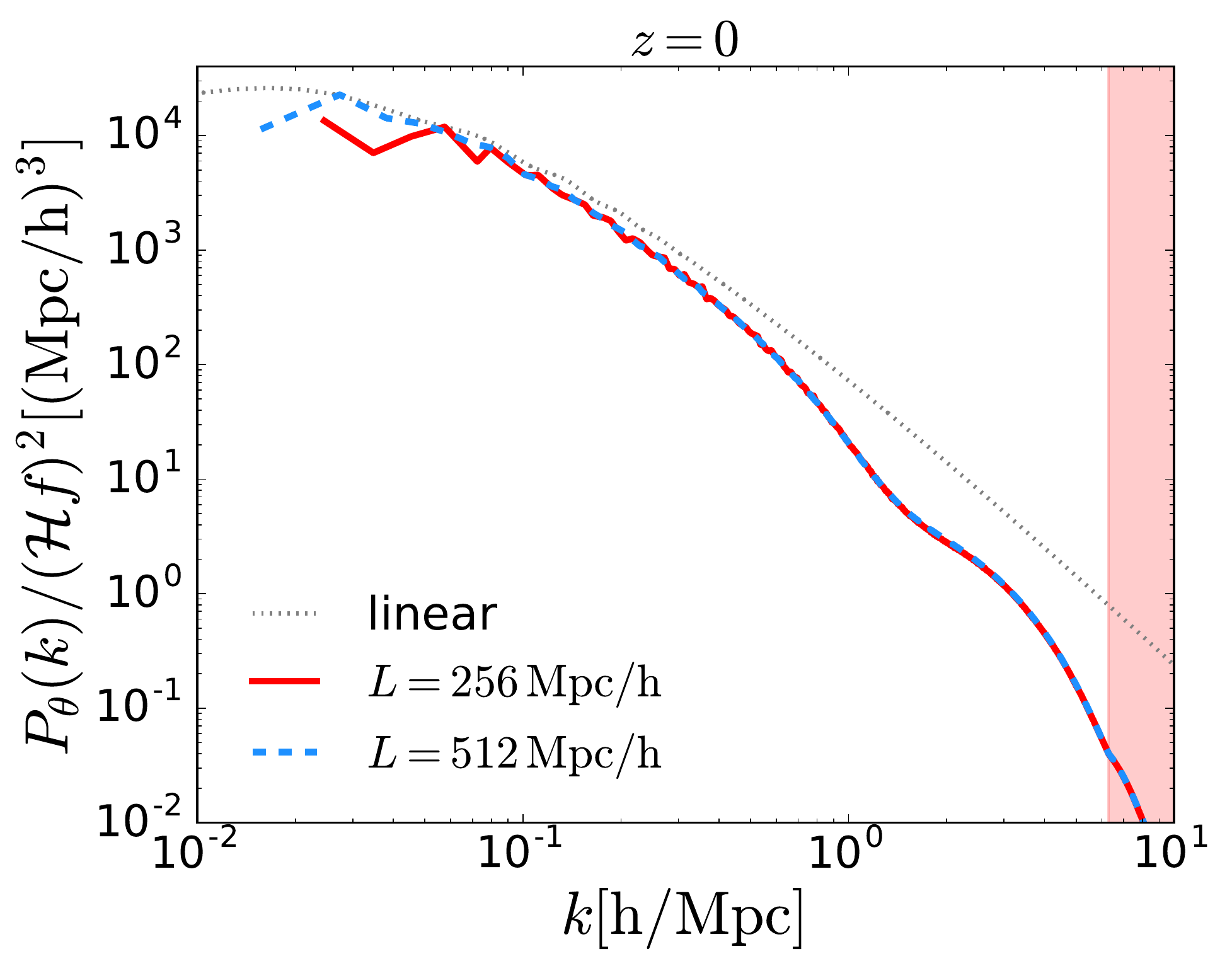}
\label{finbox:th0}
\end{subfigure}
\begin{subfigure}[b]{0.495\textwidth}
\includegraphics[width=\textwidth]{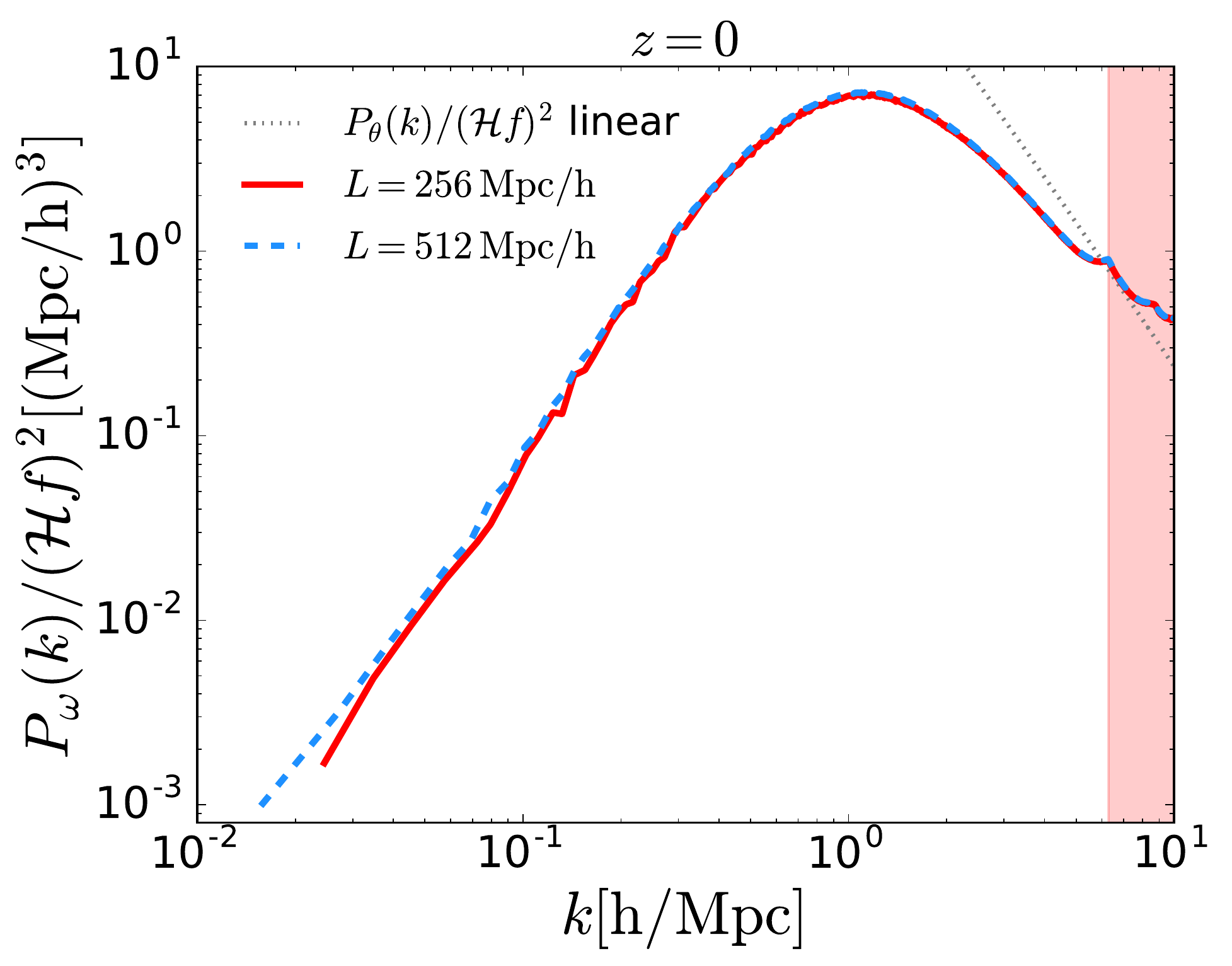}
\label{finbox:w0}
\end{subfigure}
\\
\begin{subfigure}[b]{0.495\textwidth}
\includegraphics[width=\textwidth]{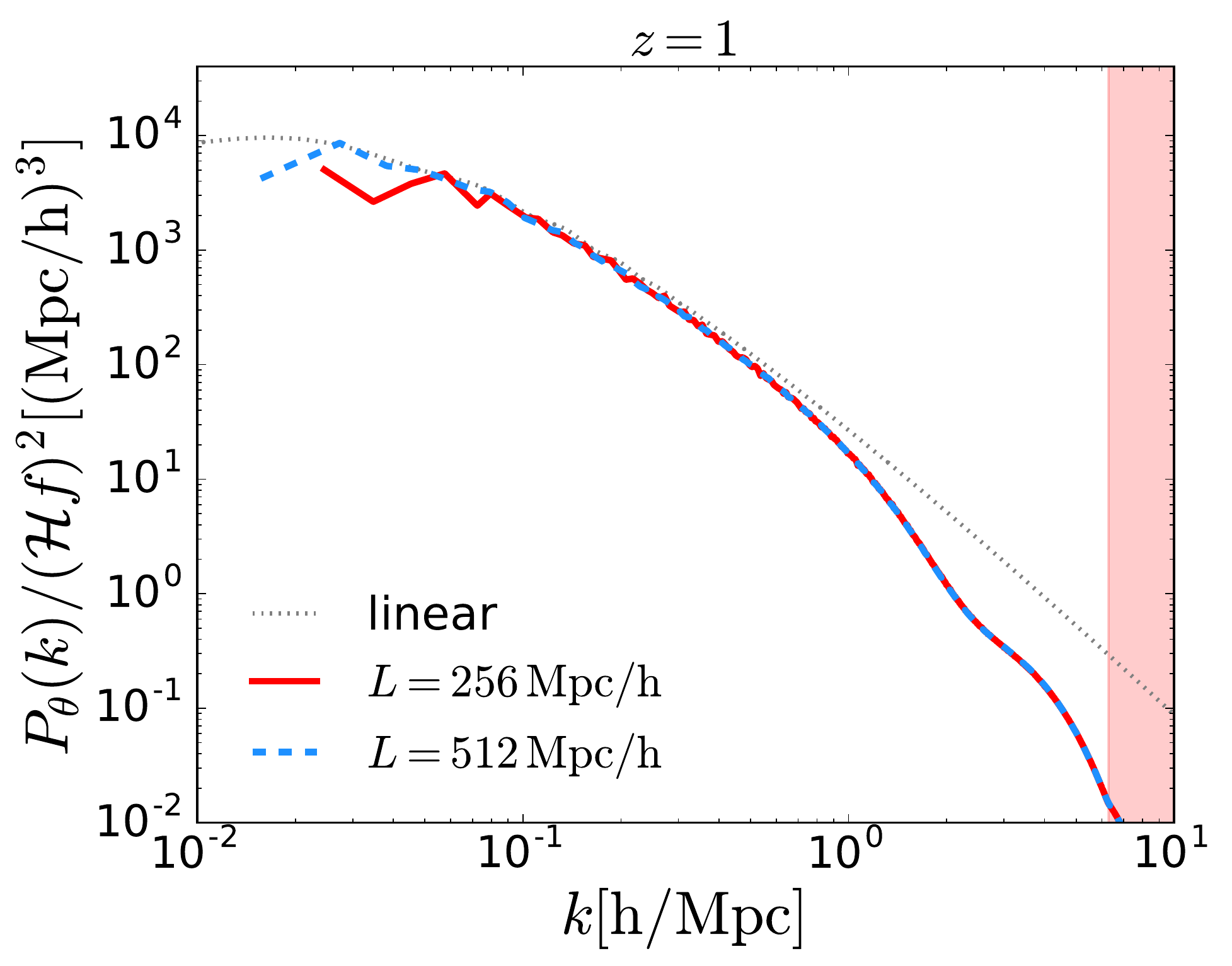}
\caption{Power spectrum - $\theta$}
\label{finbox:th1}
\end{subfigure}
\begin{subfigure}[b]{0.495\textwidth}
\includegraphics[width=\textwidth]{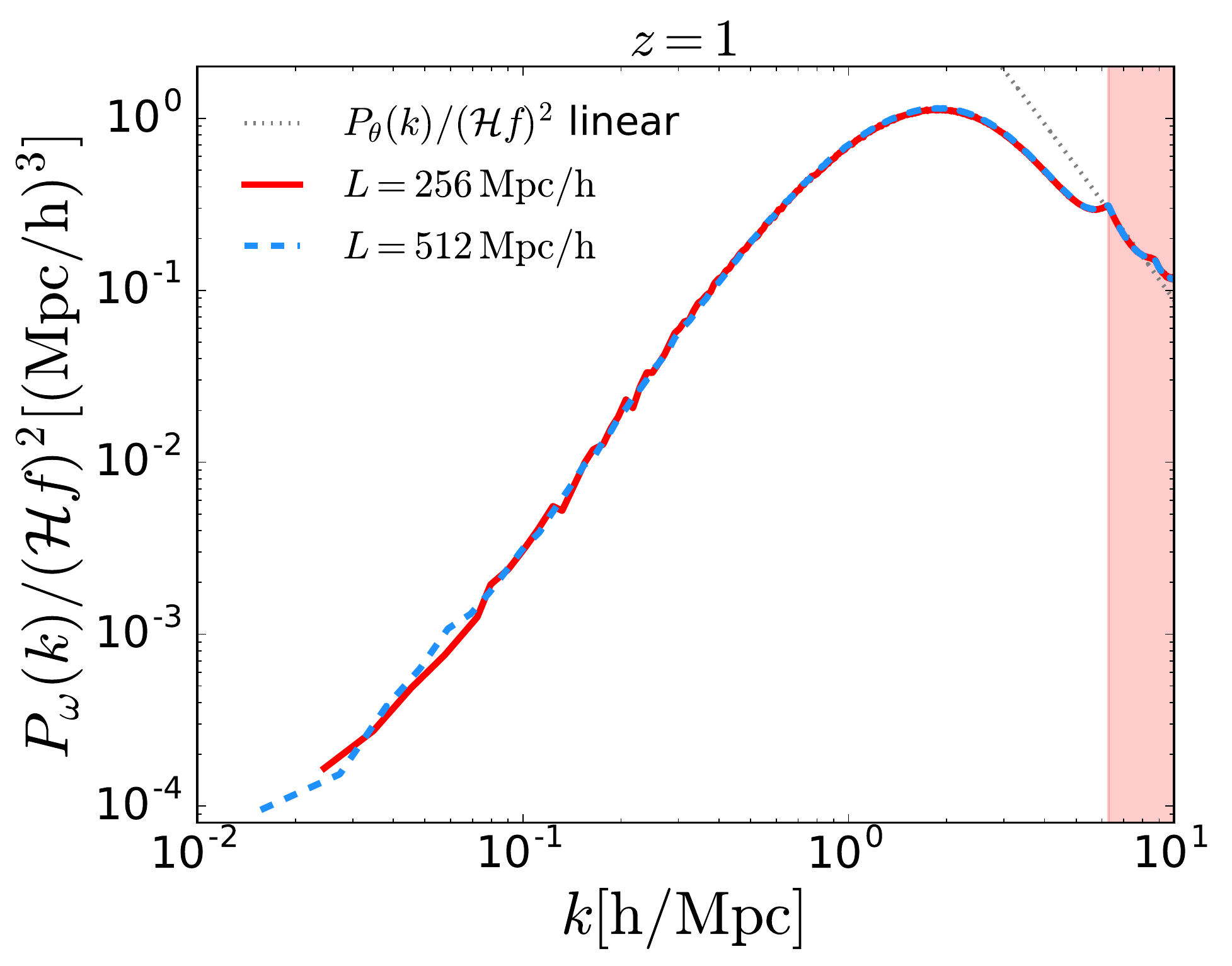}
\caption{Power spectrum - $\omega$}
\label{finbox:w1}
\end{subfigure}
\caption{Power spectrum of the divergence $\theta$ (left panel) and the vorticity $\omega$ (right panel) of the velocity field at $z = 0$ (top panels) and $z = 1$ (bottom panels) for the simulations SIM1 and BIGSIM1 in table \ref{Table-sims3}. 
The red shaded region represents the modes beyond the Nyquist frequency.}
\label{f:finbox}
\end{figure}

In figure \ref{f:finbox} we compare the power spectra for the two simulations.
Even if the BIGSIM1 is able to test a largest range of scales with respect to SIM1, we see 
that the power spectra of the two simulations agree in the common range of scales.
Therefore, we conclude that a comoving box size of $L = 256$ Mpc/$h$ is sufficiently large for 
the estimation of the divergence and vorticity power spectra on subhorizon scales.

\section{Comparison with the Delaunay Tessellation Field Estimator method}
\label{a:dtfe}
In this section we compare our method for computing the vorticity power spectrum with a standard method used in the literature to extract the velocity field
from N-body simulations, the Delaunay Tessellation Field Estimator (DTFE) method \cite{Schaap:2000se, vandeWeygaert:2007ze}. 

\begin{table}[h!]
\centering
\begin{tabular}{|c|c|c|c|c|}
\hline
\multicolumn{1}{|c|}{}                                           & 
 \multicolumn{1}{c|}{$N_\text{grid}$}                   &
  \multicolumn{1}{c|}{$N_\mathrm{part}$}                      &
  \multicolumn{1}{c|}{\textit{L} [Mpc/\textit{h}]}                        &
  \multicolumn{1}{c|}{Grid-Cell size [Mpc/\textit{h}]}                         \\
\hline
SIM1            &    256    &  512       &    256      &   1     \\
SIM2            &    512    &  1024     &    256      &   0.5      \\
\hline
\end{tabular}
\caption{
Summary of the simulations used in section \ref{a:dtfe}.
}
\label{Table-sims2}
\end{table}

The parameters of the simulations used for the comparison are given in table \ref{Table-sims2}.
The vorticity power spectrum is computed in \emph{gevolution} with the \emph{rescaled} method described in the previous sections. This is our reference method that we compare to the DTFE method.
A snapshot of the simulation is then post-processed with the DTFE public code \cite{Cautun:2011gf} in order to obtain the vorticity field from the Delaunay tessellation. 
The DTFE method employs a linear interpolation scheme in order to estimate the requested fields in each point of a regular grid, once the values of the field is known in the vertices of the Delaunay tetrahedron in which the point is located.
In the case of the vorticity field, the code computes the gradient of the velocity field, from which both vorticity and velocity divergence can be easily extracted.
The DTFE code \cite{Cautun:2011gf} employs a refined method, i.e. the fields are interpolated in $N_\text{sample}$ points inside the tetrahedron and their 
values in the grid point is computed as the volume-weighted average of the field inside the tetrahedron.
For our simulations we used $N_\text{sample} = 100$, which is large enough for the vorticity field to reach convergence.

\begin{figure}[h!]
\begin{center}
\includegraphics[width=0.6\textwidth]{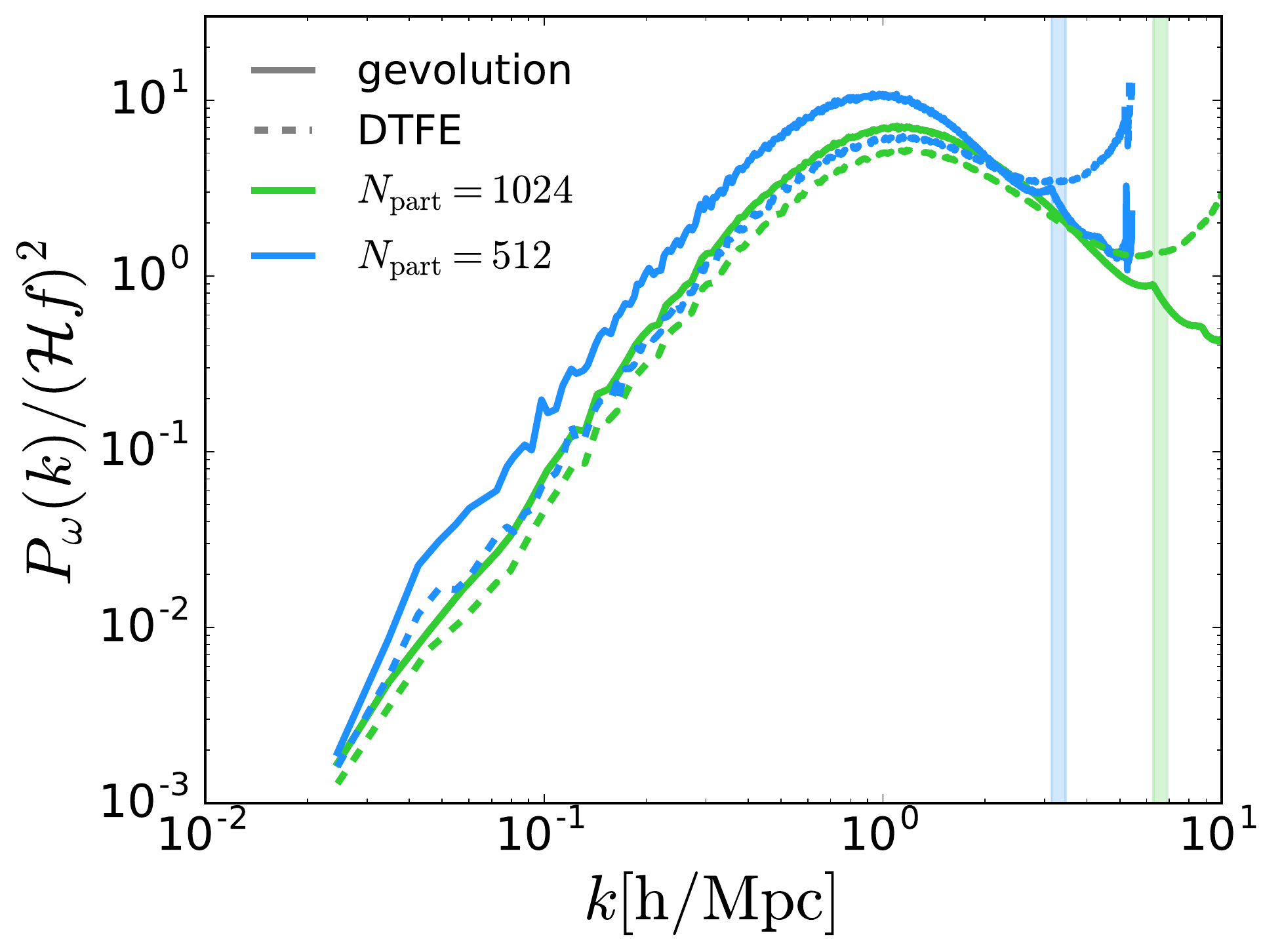}
\caption{
Comparison of the vorticity power spectrum estimated with our \emph{rescaled} method implemented in \emph{gevolution} (solid lines) and the Delaunay Tessellation Field Estimator (DTFE) method (dashed lines). We show the comparison for the simulations SIM1 and SIM2 in table \ref{Table-sims2} in blue and green, respectively. The power spectra are computed at $z = 0$.
The shaded regions mark the Nyquist frequency of the two simulations. 
}
\label{f:DTFE}
\end{center}
\end{figure}

In figure \ref{f:DTFE} we compare the two methods at $z = 0$, for the two simulations SIM1 and SIM2 in table \ref{Table-sims2}. 
We see that for the low resolution simulations the two methods reproduce the same slope of the power spectrum, but a slightly different amplitude.
We also notice that the difference in amplitude is smaller for the simulations with better resolution. This suggests that our method, based on the CIC projection of the momentum and density fields, introduces some extra projection effect with respect to
the tessellation method and therefore it has a slower convergence. 
However, this spurious effect decays when improving the mass and spatial resolution of the simulations. 
This is in agreement with the analysis presented in the appendix of \cite{Pueblas:2008uv}.
For the SIM2 in table \ref{Table-sims2} the difference is smaller than a factor two, therefore we expect the difference to be negligible for the high-resolution simulations discussed in section~\ref{s:converg}.

\bibliographystyle{JHEP}
{
\bibliography{refs-v2}
}
\end{document}